\documentclass[11pt]{article}
\usepackage{amsmath,amssymb,bm,epsf,epsfig,graphicx}
\input epsf.sty
\usepackage[utf8]{inputenc}
\usepackage[english]{babel}
\usepackage{amsmath}
\usepackage{latexsym}
\usepackage{amsfonts}
\usepackage{mathtools}
\usepackage{amssymb}
\usepackage{scalerel}
\usepackage{extpfeil}
\usepackage[left=3cm,right=3cm,top=3.1cm,bottom=3.1cm]{geometry}
\usepackage{cite}
\usepackage{tensor}
\usepackage{tikz}
\usepackage{tikz-cd}
\usepackage[mathscr]{euscript}
\usetikzlibrary{arrows.meta, bending}
\tikzset{C/.style={circle, minimum size=8mm,
		node contents={},
		append after command={\pgfextra{%
				\draw[-{Straight Barb[flex']}](\tikzlastnode.150) arc (450:110:2.8mm);}
	}}
}

\usepackage{hyperref}
\numberwithin{equation}{section}

\def\ket#1{|#1 \rangle}
\def\aver#1{\left\langle\, #1 \,\right\rangle}

\def\p{\partial}

\def \be {\begin{eqnarray}}
\def \ee {\end{eqnarray}}
\def \bal {\begin{align}}
\def \eal {\end{align}}
\def \bdm {\begin{displaymath}}
\def \edm {\end{displaymath}}
\def \ot{\otimes}

\def \tr{{\rm tr}}

\def\del {\partial}
\def\0{\nonumber}

\def\wc{\omega_\text{c}}
\def\wo{\omega_\text{o}}
\def\lc{\lambda_\text{c}}
\def\lb{\lambda_\text{b}}
\def\lo{\lambda_\text{o}}
\def\ec{\epsilon_\text{c}}
\def\eo{\epsilon_\text{o}}

\newcommand{\V}[0]{\mathbb{V}}

\begin{document}
	\begingroup\allowdisplaybreaks

\vspace*{1.1cm}

\centerline{\Large \bf The classical cosmological constant of} 

\vspace{.3cm}

 \centerline{\Large \bf  open-closed string field theory}\vspace{.3cm}
\vspace*{.1cm}

\begin{center}

{\large Carlo Maccaferri$^{(a)}$\footnote{Email: maccafer at gmail.com} and Jakub Vo\v{s}mera$^{(b)}$\footnote{Email: jvosmera at phys.ethz.ch} }
\vskip 1 cm
$^{(a)}${\it Dipartimento di Fisica, Universit\`a di Torino, \\INFN  Sezione di Torino and Arnold-Regge Center\\
Via Pietro Giuria 1, I-10125 Torino, Italy}
\vskip .5 cm
$^{(b)}${\it Institut f\"{u}r Theoretische Physik, ETH Z\"{u}rich\\
	Wolfgang-Pauli-Straße 27, 8093 Z\"{u}rich, Switzerland}

\end{center}

\vspace*{6.0ex}

\centerline{\bf Abstract}
\bigskip
We consider deformations of D-brane systems induced by a change in the closed string background in the framework of bosonic open-closed string field theory, where it is possible to unambiguously tame infrared divergences originating from both open and closed string degenerations. A closed string classical solution induces a tadpole for the open strings which shifts the open string vacuum and generates a cosmological constant  composed of two terms: one which is directly related to the closed string solution and the other which depends on the open string vacuum shift. We show that only the sum of these two terms is invariant under closed SFT gauge transformations and therefore is an observable. We conjecture that this observable is universally proportional to the shift in the world-sheet disk partition function between the starting D-brane in undeformed background and the final D-brane in deformed background, which typically includes also a change in the string coupling constant. We test the conjecture by considering a perturbative closed string solution describing deformations of a Narain compactification and, from the SFT cosmological constant, we reproduce the expected shift in the $g$-function of various D-branes living in the compactification.  In doing this we are also able to identify a surprising change in the string coupling constant at second order in the deformation.
\baselineskip=16pt
\newpage
\tableofcontents

\section{Introduction and summary}\label{sec:1}

Open and closed strings are a complete set of degrees of freedom which fully describe string theory in a perturbative phase, like quarks and gluons  in perturbative QCD. We thus expect that a full non-perturbative definition of string theory should be provided by the space-time path integral of open plus closed strings, which in turn requires a quantum field theory description, a string field theory (SFT).

Over the years, we have understood many issues regarding the non-perturbative landscape of SFT classical solutions \cite{Erler:2022agw, Erler:2019fye, Vosmera:2019mzw,  Erler:2019nmz, Erler:2019xof, Kudrna:2018mxa, Larocca:2017pbo, Kojita:2016jwe,  Kudrna:2016ack,  Maccaferri:2015cha,  Erler:2014eqa, Maccaferri:2014cpa,Kudrna:2014rya,Erler:2013wda,  Kudrna:2012re,Erler:2012qn,  Murata:2011ep, Hata:2011ke, Kiermaier:2010cf, Bonora:2010hi, Schnabl:2005gv} and the explicit constructions of complete RNS actions \cite{Moosavian:2019ydz, Kunitomo:2019glq, Erler:2017onq, Konopka:2016grr, Erler:2016ybs, Sen:2015uaa,  Kunitomo:2015usa}. More recently SFT methods have been used to address non-perturbative D-instantons corrections to closed string scattering amplitudes \cite{Chakravarty:2022cgj, Eniceicu:2022dru,
Eniceicu:2022nay, Alexandrov:2022mmy,  Agmon:2022vdj,
Alexandrov:2021dyl, Alexandrov:2021shf, Sen:2021jbr,
Sen:2021qdk, Sen:2020eck, Sen:2020ruy, Sen:2020oqr, Sen:2020cef, Sen:2019qqg, Sen:2019jpm, Sen:2016qap}, which also provided a first-principle resolution of several world-sheet drawbacks associated with degenerations of Riemann surfaces. See \cite{Erler:2019loq,Erler:2019vhl,deLacroix:2017lif} for recent reviews on SFT. 

In a recent series of works\cite{Maccaferri:2021ksp, Maccaferri:2021lau, Maccaferri:2021ulf,Erbin:2020eyc, Koyama:2020qfb}, with the aim of understanding open/closed dynamics in the rigorous framework of string field theory, there has been interest in considering open (super)string field theories based on Witten star product with the addition of a gauge invariant coupling to on-shell closed string states (the so-called Ellwood Invariant, \cite{Hashimoto:2001sm, Gaiotto:2001ji, Ellwood:2008jh})  both in bosonic and in Type II theories. The use of Witten vertex is advantageous  because it provides the simplest decomposition of the bosonic worldsheet moduli space in terms of open string propagators, cubic vertices and elementary open-closed couplings\cite{Zwiebach:1992bw, Zwiebach:1990az}. In the above-mentioned works it has been shown  that, although the microscopic open/closed coupling is just linear in the physical closed string field, integrating out classically the massive open string modes generates an effective action containing all possible disk amplitudes involving the deforming physical closed strings and the massless open strings. In general these amplitudes can contain infrared divergences related to regions of open or closed string degenerations and the advantage of the SFT formulation is that it gives a way to unambiguously tame these divergences \cite{Sen:2016qap,  Larocca:2017pbo, Sen:2019jpm}. However, in the very economic formulation we just described (where the only dynamical variables are open strings) there is no natural way to tame IR divergences originating from closed string degenerations, because there are only open string propagators. As a matter of fact this is a problem only when closed strings approach the zero momentum limit or other special kinematical configurations where on shell states would propagate inside the amplitude. In the other cases one can compute closed string amplitudes at generic external momenta where the amplitude is finite and then analytically continue the result to the given (non vanishing) closed string momenta. This scenario  includes the recent explorations of D-instanton corrections to closed strings amplitudes which are affected by intrinsic IR divergences from open string degenerations  but which can be made finite in the region of closed string degeneration. 

 There are however other scenarios of open/closed dynamics where it  is fundamental to understand the soft behaviour of amplitudes involving external closed string states. In general these ``soft'' open-closed amplitudes occur whenever one tries to understand how a D-brane system responds to a change in the closed string background, which will typically be described by long-wave spacetime deformations corresponding to very soft closed string vertex operators. In \cite{Maccaferri:2021ksp} we have identified the simplest of such amplitudes as a bulk two-point function on the disk of  zero-momentum physical closed string states. This term, which is at second order in the closed string deformation, is part of a cosmological constant which is generated by the shift of the open string field as a response to the tadpole given by the elementary open-closed coupling. Although this term is non-dynamical for open strings on the shifted vacuum, it defines a vacuum energy and therefore it is important to understand its physical meaning. However, before even speculating about the  nature of this term,  it is necessary to be able to actually compute it, {\it i.e.} to establish whether the quantity is well-defined or not. Since this term corresponds to closed string amplitudes on a disk, there are in general divergences arising from closed strings hitting the boundary (open string degeneration) and closed strings hitting other closed strings in the bulk (closed string degeneration). As alluded above, while the open string propagator gives a perfectly consistent way to deal with open string degeneration, there does not seem to be a natural prescription to tame bulk collisions without explicitly (or implicitly, as done in \cite{Maccaferri:2021ksp}) introducing closed string propagators in the amplitudes. Therefore the physical understanding of the cosmological constant that is generated by a D-brane system after a deformation in the closed string background is something that can be fully understood only in the framework of open-closed string field theory.
 
Covariant open-closed string field theory has been formulated by Zwiebach \cite{Zwiebach:1990qj, Zwiebach:1997fe} as an intrinsically quantum theory which provides a solution to the quantum BV master equation on the space of open plus closed strings. Out of this theory one can extract two classical BV master actions given respectively by the truncation to only closed strings and genus zero vertices with no boundary (giving rise to an $L_\infty$ algebra) or to only open strings with genus zero  and one-boundary vertices (giving rise to an $A_\infty$ algebra). Interestingly, there is no way to get a purely classical theory with propagating open and closed strings: for example a one-loop open string tadpole diagram containing   a ``quantum'' open string annulus and another diagram given by  a ``classical''  closed string propagator connecting two boundary states (one of which with a boundary insertion) will cover different moduli space regions of the same physical amplitude. There is however a consistent classical truncation that is useful to study the open string dynamics in a deformed closed string background given by a closed string classical solution, as observed years ago  \cite{Zwiebach:1997fe}. The starting point for this theory is to take  only the  genus zero vertices with zero or one boundary, with all possible boundary and bulk punctures. These vertices obey a set of homotopy relations which we call {\it sphere-disk homotopy algebra} (SDHA). This homotopy algebra is in turn  an extension of the {\it open-closed homotopy algebra}  of Kajiura and Stasheff \cite{Kajiura:2004xu, Kajiura:2005sn}, (commonly referred to as OCHA) where disk couplings always include at least one open string. On top of the OCHA vertices, the SDHA thus contains also disks with only bulk punctures which do not give rise to dynamical terms for open strings. Nevertheless, as we will see in detail, these couplings are needed for overall consistency and our results in this paper are  based on the part of the SDHA which is not  contained in the OCHA of Kajiura and Stasheff. It is only using this extended set of vertices that one can  consistently construct an open string field theory in a deformed background by  freezing the closed string field to a closed string field theory  classical solution and letting the open string field free to fluctuate. This is so because, after having replaced the closed string field theory solution in the open-closed theory, we are left with an open string field theory characterized by a weak $A_\infty$ structure with a non-vanishing zero-product, a tadpole. Therefore we have to shift the open string field to go to a locally gaussian point and in doing this we generate a vacuum energy. This vacuum energy coming from the shift of the open string field is not invariant under a gauge transformation of the closed string solution. This is clearly a problem which luckily is solved by combining this open string vacuum energy with the SDHA purely closed string vertices evaluated on the classical solution. The sphere part of these vertices is the classical action of the classical solution in closed string field theory and it is expected to vanish, up to possible boundary contributions, as recently argued in \cite{Erler:2022agw}. The disk part, instead, is not vanishing and, like the open string vacuum energy, is also not gauge invariant. To recover a gauge invariant quantity, we have to sum the vacuum energy from the open string vacuum shift with the vacuum energy coming from the purely closed string disk couplings. Indeed, using the open-closed homotopy relations and the closed string field theory equation of motion, we will show that only
the sum of the two terms is gauge invariant under the $L_\infty$ gauge transformation of the closed string classical solution. Therefore this combined open+closed quantity, the full cosmological constant, is an observable that should have a physical meaning. This is the first main result of this paper.

 We  conjecture that this cosmological constant is proportional to the shift in the disk partition function between the starting D-brane in the undeformed background and the final D-brane in the deformed background. To be precise,  let us consider an initial closed string background  CFT$_{0}$ with coupling constant $g_s^{(0)}$ and an open string background given by a (matter) boundary state $\|B_0\rangle\!\rangle$. Let then $\Phi_{\rm cl}$ be a closed string classical solution describing a new closed string background $({\rm CFT}_*,\,g_s^{(*)})$ (with possibly a new string coupling constant\footnote{The initial and final coupling constants can differ for example because the closed string solution turns on the ghost-dilaton which is known to change the coupling constant to leading order without touching the background CFT, \cite{Bergman:1994qq}. Interestingly, as we will see, the string coupling constant can also change (to subleading order) even if we don't turn on the ghost-dilaton.}). Let $\|B_*\rangle\!\rangle$ be the (matter) boundary state of the new D-brane system in the new closed string background, described by the open-string vacuum-shift solution $\Psi_\mathrm{v}$ (which is a classical solution of the weak $A_\infty$ equations of motion), and finally, let $\Lambda(\Phi_{\rm cl};\Psi_\mathrm{v})$ be the gauge-invariant cosmological constant that is generated by the classical closed string solution $\Phi_\mathrm{cl}$, together with the open string vacuum shift solution $\Psi_\mathrm{v}$. Then our conjecture reads
\begin{equation}
-\frac{1}{2\pi^2 g_s^{(*)}}\langle 0\| B_*\rangle\!\rangle=-\frac{1}{2\pi^2 g_s^{(0)}}\langle 0 \| B_0\rangle\!\rangle + \Lambda(\Phi_{\rm cl};\Psi_\mathrm{v})\,,\label{conj}
\end{equation}
where we normalize the interaction vertices so that they do not contain explicit powers of $g_s^{(0)}$. A couple of comments are in order. First, since the combination $(-2\pi^2 g_s)^{-1}\langle 0 \| B\rangle\!\rangle $ is universally proportional to the disk partition function $Z_\mathrm{disk}$ of the given open-closed background (setting $\alpha^{\prime}=1$), then \eqref{conj} says that the cosmological constant $\Lambda(\Phi_\mathrm{cl};\Psi_\mathrm{v})$ indeed measures the shift of the disk partition function when changing the background from the perturbative vacuum $(\mathrm{CFT}_0,\, g_s^{(0)},\|B_0\rangle\!\rangle)$ to the new open-closed background $(\mathrm{CFT}_*,\, g_s^{(*)},\|B_*\rangle\!\rangle)$ given by $\Phi_\mathrm{cl}$ and  $\Psi_\mathrm{v}$.
We also emphasize that the closed string coupling $g_s^{(*)}$ of the shifted background is sensitive only to $\Phi_\mathrm{cl}$ and is universal with respect to the open-string vacuum shift solution $\Psi_\mathrm{v}$.
Finally, notice that this is a generalization of Sen's conjecture \cite{Sen:1999xm} to the case when both open and closed string backgrounds are allowed to dynamically change. The old Sen's conjecture is then recovered by setting $\Phi_\mathrm{cl}=0$ and taking $\Psi_\mathrm{v}$ to be any solution to the classical OSFT equations of motion.

In the remaining part of the paper we test this conjecture in a simple setting where explicit worldsheet calculations are possible: we consider a perturbative closed string solution which describes a continuous deformation of the starting background. At second order in the closed string deformation, the cosmological constant is then given by a sum of two Feynman diagrams which assemble to reproduce a disk two-point amplitude. 
In general such amplitude exhibit potential divergences from open or closed string degenerations, but we can simply deal with them using the open or closed string propagators respectively. To be concrete, we focus on a Narain compactification on a $d$-dimensional torus with a metric $g_{ij}$ and a Kalb-Ramond field $B_{ij}$,
combined in $E_{ij}=(g+B)_{ij}$. Provided that we fix a particular D-brane wrapping the Narain lattice, we can use the corresponding boundary state $\|B(E)\rangle\!\rangle$ to extract its $g$-function $\langle 0\|B(E)\rangle\!\rangle$ as a function of the Narain moduli $E$ and investigate how the $g$-function changes to a new value $\langle 0\| B(E+\epsilon_\sigma)\rangle\!\rangle$ as we change the CFT moduli by deforming $E$ as $ E\longrightarrow E+\epsilon_\sigma$, where $\epsilon_\sigma$ is the deformation parameter of the exact CFT sigma-model. 
On the other hand, in our SFT setting, we can  deform the closed string background $E$ by giving a vev to the exactly marginal field $\epsilon_{ij}\,c\del X^i\,\bar c \bar\del X^j$, where the SFT modulus $\epsilon$ is generally  non-trivially related to the sigma-model modulus $\epsilon_\sigma$. This generates a closed string field theory solution $\Phi_{\rm cl}(\epsilon)$ which we build explicitly up to second order in $\epsilon$. To the same order, we can construct a perturbative open string vacuum shift solution $\Psi_\mathrm{v}(\epsilon)$ which corresponds to the starting D-brane adapting itself to the new closed string background. Finally, we can compute the gauge invariant cosmological constant $\Lambda(\epsilon)\equiv\Lambda(\Phi_{\rm cl}(\epsilon);\Psi_\mathrm{v}(\epsilon))$. In order to compare between the disk partition function shifted by the SFT cosmological constant $\Lambda(\epsilon)$ and the deformed CFT $g$-function $\langle 0\|B(E+\epsilon_\sigma)\rangle\!\rangle$,
we finally need to fix the relation between $\epsilon_\sigma$ and $\epsilon$. This is a question which is completely independent of boundary conditions and which can be perturbatively answered in closed string field theory by matching the spectrum of fluctuations on the background given by the classical solution, with the spectrum of the exact CFT at background value $E+\epsilon_\sigma$. We are able to find this relation up to second order in the limit of large closed-string stub by adapting an analogous computation by Sen \cite{Sen:2019jpm}
\begin{equation}
\epsilon_{\sigma}=\epsilon+\frac12 \epsilon\,g^{-1}\,\epsilon+\mathcal{O}(\epsilon^3)\,.\label{eq:12}
\end{equation}
Given all this input, we are then finally able to show that the sum of the initial disk partition function and the SFT cosmological constant $\Lambda(\epsilon)$ factorizes into the shifted $g$-function $\langle 0 \| B(E+\epsilon_\sigma)\rangle\!\rangle$ times a factor,  independent of the particular conformal boundary state $\|B(E)\rangle\!\rangle$ which we choose as a starting open string background.  Explicitly,  up to second order in $\epsilon$, we find 
\begin{align}
-\frac{1}{2\pi^2 g_s^{(0)}}\langle 0 \| B(E)\rangle\!\rangle + \Lambda(\epsilon)=-\frac{1}{2\pi^2 g_s^{(\ast)}(\epsilon)}\langle 0 \| B(E+\epsilon_\sigma(\epsilon))\rangle\!\rangle \,,
\end{align}
where the universal factor $g_s^{(\ast)}(\epsilon)$ is given by
\begin{align}
g_s^{(*)}(\epsilon)=g_s^{(0)}\bigg(1+\frac1{32}\tr[\epsilon g^{-1} \epsilon^T g^{-1}]+\mathcal{O}(\epsilon^3)\bigg)\label{gdef}
\end{align}
which, being independent of the D-branes boundary conditions, is naturally identified with a (subleading) shift in the string coupling constant. 
This is the second main result of this paper.

The paper is organized as follows. In section \ref{sec:2} we review the structure of open-closed bosonic string field theory. In particular, in subsection \ref{sec:2.1} we describe the homotopy algebra (SDHA) that is obtained by truncating the full quantum BV master equation to the level of spheres and disks. We then derive the relations of SDHA from the BV master equation in subsection \ref{sec:2.2}. Then in subsection \ref{sec:2.3} we construct a deformed open string field theory by freezing the closed string field to a generic classical solution and we eliminate  the open string tadpole by expanding the open string field around an open string vacuum shift solution. In subsection \ref{sec:2.4} we show that the obtained cosmological constant, apart from being manifestly invariant under the $A_\infty$ gauge transformation of the open-string vacuum shift solution, is also gauge invariant under the $L_\infty$ gauge transformation of the closed string classical solution. Then we formulate our conjecture \eqref{conj}. In sections \ref{sec:3} and \ref{sec:4}, we test the conjecture perturbatively. In particular, in subsections \ref{sec:3.1}, \ref{sec:3.2} and \ref{sec:3.3}, we compute the cosmological constant induced by an arbitrary marginal deformation of the closed string background (to which a given D-brane system is able to adapt) up to quadratic order in the modulus, verifying that it is given by a collection of tree-level disk amplitudes. In subsections \ref{sec:3.4} and \ref{sec:3.5}, we then derive a concrete expression for the cosmological constant in such a setting in terms of matter CFT correlators. Finally, in subsections \ref{sec:4.1} and \ref{sec:4.2}, we fix a particular closed-string background with moduli ($d$-dimensional Narain lattice with various conformal boundary conditions) and test the conjecture \eqref{conj} by considering the classical closed SFT solution which implements the change of the moduli. In section \ref{sec:5} we conclude with an outlook of the next steps of our program. Three appendices contain important material which is needed in the main text. In appendix \ref{app:A} we construct the $\mathrm{SL}(2,\mathbb{C})$ open-closed vertices which are needed for our explicit computation and we fix the various free parameters to make sure that the moduli space of the two punctured disk is covered with just two diagrams, by checking the corresponding SDHA relation. In appendix \ref{app:B}, we review the construction of the BRST cohomology around a classical SFT solution corresponding to a marginal deformation, giving all-order expressions using the homotopy transfer. Finally, in appendix \ref{app:C}, we relate the SFT deformation parameter to the sigma model modulus by studying the spectrum of fluctuations of a closed string field theory solution for a marginal deformation.
\section{Background shifts in open-closed string field theory}\label{sec:2}

To construct an open-closed string field theory \cite{Zwiebach:1990qj, Zwiebach:1997fe} we need to pick a closed string background CFT$_0$, a string coupling constant $g_s$ and a compatible open string background BCFT$_0$.  
Let $\Phi=\phi^i\ket{c_i}\in {\cal H}^\text{c}$ be the Batalin-Vilkovisky (BV) closed string field which is a generic bulk state obeying
\be
{\cal H}^\text{c}:\quad b_0^-\ket{\Phi}=L_0^-\ket{\Phi}=0\,,
\ee
with an overall vanishing degree which is the sum of the BV degree of the spacetime polarizations $\phi^i$ and of the basis states $\ket{c_i}$ whose degree is taken to be the ghost number minus two.
At the same time we take $\Psi=\psi^i\ket{o_i}\in{\cal H}^\text{o}$ to be the BV open string field which is a boundary operator with total vanishing degree, where the basis states $\ket{o_i}$ are assigned to have a degree given by the ghost number minus one.
The BV action functional has a natural topological expansion in the genus $g$ and the number of boundaries $b$ 
\be
S(\Phi,\Psi)=\sum_{g=0}^\infty\sum_{b=0}^\infty \, S_{g,b}(\Phi,\Psi)\,,
\ee
where 
\be
S_{g,b}(\Phi,\Psi)=-g_s^{2g-2+b}\sum_{k,l_i}{\cal V}_{g,b}^{k,\{l_i\}}(\Phi^{\wedge k};\Psi^{\otimes l_1},\cdots,\Psi^{\otimes l_b})
\ee
is a sum of BCFT$_0$ correlators on corresponding Riemann surfaces of the given number of bulk and boundary insertions (at every boundary) with, where needed,  an integration on subregions of moduli space accompanied by $b$-ghost insertions. The wedge $\wedge$ is the (graded) symmetrized tensor product. The consistency of this construction relies on the fact that the full action functional should obey the quantum BV master equation
\be
(S,S)+2\Delta S=0\,,
\ee
where we define the open-closed BV bracket $(\cdot,\cdot)$ in terms of the closed- and open-string BV brackets as
\begin{align}
    (\cdot,\cdot) = (\cdot,\cdot)_\text{c}+(\cdot,\cdot)_\text{o}\,.
\end{align}
Denoting $\phi^i$ and $\psi^i$ the dual basis states of $\mathcal{H}^\text{c}$  and $\mathcal{H}^\text{o}$, respectively, (i.e. the BV spacetime fields) we then define
\begin{subequations}
\begin{align}
    (\cdot,\cdot)_\text{c} &= g_s^{2}\frac{\overset{\leftarrow}{\p}}{\p\phi^i}\omega_\text{c}^{ij}\frac{\overset{\rightarrow}{\p}}{\p\phi^j}\,,\\
    (\cdot,\cdot)_\text{o} &= g_s\frac{\overset{\leftarrow}{\p}}{\p\psi^i}\omega_\text{o}^{ij}\frac{\overset{\rightarrow}{\p}}{\p\psi^j}\,,
\end{align}
\end{subequations}
where 
\begin{subequations}
\begin{align}
&\omega_\text{c}:(\mathcal{H}^\text{c})^{\otimes 2}\longrightarrow \mathbb{C}\,,\\ 
&\omega_\text{o}:(\mathcal{H}^\text{o})^{\otimes 2}\longrightarrow \mathbb{C}    \,,
\end{align}
\end{subequations}
are the closed and open symplectic forms, which are related to the standard BPZ inner product as
\begin{subequations}
\begin{align}
\omega_\text{c}(\Phi_1,\Phi_2)&=(-1)^{d(\Phi_1)}\aver{\Phi_1,c_0^-\Phi_2}_\text{c}\,,\\
\omega_\text{o}(\Psi_1,\Psi_2)&=(-1)^{d(\Psi_1)}\aver{\Psi_1,\Psi_2}_\text{o}\,,
\end{align}
\end{subequations}
where $d$ denotes the degree. Then $\omega^{ij}$ is the inverse of $\omega_{ij}\equiv\omega(e_i,e_j)$, where $e_i$ is a basis of the corresponding open or closed Hilbert spaces.
Similarly for the open-closed BV Laplacian $\Delta$ we write
\begin{align}
    \Delta = \Delta_\text{c}+\Delta_\text{o}\,,
\end{align}
which are defined as
\begin{subequations}
\begin{align}
    \Delta_\text{c} &= \frac{1}{2}g_s^{2}(-1)^{d(\phi^i)}\omega_\text{c}^{ij}\frac{\overset{\rightarrow}{\p}}{\p \phi^i}\frac{\overset{\rightarrow}{\p}}{\p \phi^j}\,,\\
    \Delta_\text{o} &= \frac{1}{2}g_s(-1)^{d(\psi^i)}\omega_\text{o}^{ij}\frac{\overset{\rightarrow}{\p}}{\p \psi^i}\frac{\overset{\rightarrow}{\p}}{\p \psi^j}\,.\label {deltao}
\end{align}
\end{subequations}

\subsection{Tree-level truncation of  open-closed string field theory}\label{sec:2.1}

If we denote $\Sigma_{g,b}$ the space of all possible correlators on a Riemann surface at genus $g$ with $b$ boundaries, we can easily see that the basic BV structures operate at the following level
\begin{subequations}
\begin{align}
(S_{g_1,b_1},S_{g_2,b_2})_\text{c}&\in\Sigma_{g_1+g_2,b_1+b_2}\,,\\
(S_{g_1,b_1},S_{g_2,b_2})_\text{o}&\in\Sigma_{g_1+g_2,b_1+b_2-1}\,,\\
\Delta_\text{c} S_{g,b}&\in\Sigma_{g+1,b}\,,\\
\Delta^{(1)}_\text{o} S_{g,b}&\in\Sigma_{g,b+1}\,,\\
\Delta^{(2)}_\text{o} S_{g,b}&\in\Sigma_{g+1,b-1}\,, \quad {b\geq 2}
\end{align} 
\end{subequations}
where we have decomposed the open string laplacian as $\Delta_\text{o}=\Delta^{(1)}_\text{o}+\Delta^{(2)}_\text{o}$, distinguishing the cases in \eqref{deltao} in which the two open strings selected by the derivatives sit on the same boundary ($\Delta^{(1)}_\text{o}$) or on different boundaries of the same surface ($\Delta^{(2)}_\text{o}$).
This allows to topologically decompose the full BV master equation to surfaces at given genus and boundary number.
In particular we notice \cite{Zwiebach:1997fe} that the $g=0$ part of the full action, $S_0\equiv \sum_{b=0}^{\infty} S_{0,b}$, obeys the master equation
\be
(S_0,S_0)_\text{c}+(S_0,S_0)_\text{o}+2\Delta^{(1)}_\text{o} S_0=0\,.\label{zeroBV}
\ee
In this $g=0$ theory (where closed strings are treated at tree-level) we would like to see whether also open strings can be restricted to be classical. It is however well-known that this is not possible. Decomposing the $(g=0)$ master equation \eqref{zeroBV} in the number of boundaries $b$, we find
\begin{subequations}
\begin{align}
b=0:&\quad (S_{0,0},S_{0,0})_\text{c}=0\,,\label{b0}\\
b=1:&\quad 2(S_{0,0},S_{0,1})_\text{c}+	(S_{0,1},S_{0,1})_\text{o}=0\,,\label{b1}\\
b=2:&\quad (S_{0,1},S_{0,1})_\text{c}+2(S_{0,0},S_{0,2})_\text{c}+2(S_{0,1},S_{0,2})_\text{o}+2\Delta^{(1)}_\text{o} S_{0,1}=0\,.\\
&\vdots\nonumber
\end{align}
\end{subequations}
From here it is apparent that it is not possible to keep both closed and open strings dynamical and have an action which solves the classical BV master equation. In the presence of dynamical closed strings, open strings are necessarily quantum ($b\geq1$). Notice however that if we set the closed string field $\Phi$ to 0 then we get a classical open string field theory action $S_{0,1}(0,\Psi)$ which solves the classical BV master equation, because
\be
(S_{0,1},S_{0,1})_\text{o}{\Big |}_{\Phi=0}=0\,.
\ee
This in fact generalizes when $\Phi=\Phi_{\rm cl}$ is a classical solution of $S_{0,0}(\Phi)$, which is the classical closed string field theory action. Indeed, looking at \eqref{b1}, we see that 
\be
(S_{0,0},S_{0,1})_\text{c}{\Big |}_{\Phi=\Phi_{\rm cl}}=\omega_c\left(\frac{\delta S_{0,0}}{\delta\Phi},\frac{\delta S_{0,1}}{\delta\Phi}\right){\Big |}_{\Phi=\Phi_{\rm cl}}=0\,,
\ee
since $\frac{\delta S_{0,0}}{\delta\Phi}{\Big |}_{\Phi=\Phi_{\rm cl}}=0$,
so that \eqref{b1} implies
\be
(S_{0,1},S_{0,1})_\text{o}{\Big |}_{\Phi=\Phi_{\rm cl}}=0\,.
\ee
Therefore the quantity
\be
S_{\Phi_{\rm cl}}(\Psi)\equiv  S_{0,1}(\Phi_{\rm cl},\Psi)\label{action-def}
\ee
is a solution to the open BV classical master equation
\be
(S_{\Phi_{\rm cl}},S_{\Phi_{\rm cl}})_\text{o}=0\,.\label{open-bv}
\ee

\subsection{Sphere-Disk Homotopy Algebra (SDHA)}\label{sec:2.2}
Before entering into the physics described by \eqref{action-def}, it is important to translate the BV equations \eqref{b0} and \eqref{b1} in the language of the homotopy algebras obeyed by the open-closed vertices which make up the action. Let's start by reviewing how this works for the classical closed string field theory action $ S_{0,0}(\Phi)$
\begin{align}
 S_{0,0}(\Phi)=-g_s^{-2}\sum_{k=1}^\infty \mathcal{V}_{0,0}^{k,\emptyset}(\Phi^{\wedge k})\,.
\end{align}
Let us assume the following form for the $b=0$ vertices
\begin{align}
    \mathcal{V}_{0,0}^{k+1,\emptyset}(\Phi^{\wedge (k+1)}) = \frac{1}{(k+1)!}\omega_\text{c}(\Phi,l_{k}(\Phi^{\wedge k}))\,,
\end{align}
where
\begin{align}
l_k:(\mathcal{H}^\text{c})^{\wedge k}\longrightarrow \mathcal{H}^\text{c}    
\end{align}
is a set of graded-symmetric degree $+1$ products, which are cyclic with respect to $\omega_\text{c}$, namely
\begin{align}
    \omega_\text{c}(\Phi_1,l_k(\Phi_2,\ldots,\Phi_{k+1}))&=-(-1)^{d(\Phi_1)}\omega_\text{c}(l_k(\Phi_1,\ldots,\Phi_k),\Phi_{k+1})\,.
\end{align}
Using the fact that the BV closed string field is globally degree even, we can now explicitly evaluate the classical master equation 
\begin{subequations}
\begin{align}
   g_s^{2} (S_{0,0},S_{0,0})_\text{c} &= \sum_{k,m\geqslant 1}\frac{1}{k!m!}\omega_\text{c}(l_k(\Phi^{\wedge k}),l_m(\Phi^{\wedge m} ))\\
    &= \sum_{k,m\geqslant 1}\Big(\frac{m}{k+m}+\frac{k}{k+m}\Big)\frac{1}{k!m!}\omega_\text{c}(l_k(\Phi^{\wedge k}),l_m(\Phi^{\wedge m} ))\\
    &= \sum_{k,m\geqslant 1}\frac{2k}{k+m}\frac{1}{k!m!}\omega_\text{c}(l_k(\Phi^{\wedge k}),l_m(\Phi^{\wedge m} ))\\
    &= -\sum_{k,m\geqslant 1}\frac{2}{k+m}\frac{1}{(k-1)!m!}\omega_\text{c}(\Phi,l_k(\Phi^{\wedge (k-1)},l_m(\Phi^{\wedge m} )))\\
    &= -\sum_{k,m\geqslant 1}\frac{2}{(k+m)!}{k+m-1\choose m}\omega_\text{c}(\Phi,l_k(\Phi^{\wedge (k-1)},l_m(\Phi^{\wedge m} )))\\
    &= -\sum_{r=1}^\infty\frac{2}{(r+1)!}\sum_{k=1}^r\omega_\text{c}(\Phi,l_kl_{r+1-k}(\Phi^{\wedge r}))\,.
\end{align}
\end{subequations}
Hence, we observe that we can solve the master equation $(S_{0,0},S_{0,0})_\text{c}=0$ by requiring that the products $l_k$ satisfy the $L_\infty$ relations 
\begin{align}
    \sum_{k=1}^r l_k l_{r+1-k}=0\,.
\end{align}
Let us now consider the $b=1$ part of the $g=0$ open-closed action
\begin{align}
S_{0,1}(\Phi,\Psi) &=Z_{\rm disk} -g_s^{-1}\sum_{\substack{k\geqslant 0,l\geqslant 0\\
    k+l\geqslant 1}}\mathcal{V}_{0,1}^{k,l}(\Phi^{\wedge k};\Psi^{\otimes l})\,.
\end{align}
Note that, following \cite{Zwiebach:1997fe}, we have included the disk partition function as a non-dynamical term, for later convenience. Note also that the open-string tadpole $\mathcal{V}_{0,1}^{0,1}$ is identically vanishing.
In the non-constant part of the open-closed  action we can further identify the terms involving only closed strings on the disk $\mathcal{V}_{0,1}^{k,0}$
\begin{align}
    \mathcal{V}_{0,1}^{k+1,0} (\Phi^{\wedge k+1})= \frac{1}{(k+1)!}\omega_\text{c}(\Phi,l_{k,0}(\Phi^{\wedge k}))\,,
\end{align}
where $l_{k,0}$ for $k\geqslant 0$ (not to be confused with the $l_k$ products discussed above) are maps
\begin{align}
l_{k,0}:(\mathcal{H}^\text{c})^{\wedge k}\longrightarrow \mathcal{H}^\text{c}    
\end{align}
which are again a set of graded-symmetric degree $+1$ products, which are cyclic with respect to $\omega_\text{c}$
\begin{align}
    \omega_\text{c}(\Phi_1,l_{k,0}(\Phi_2,\ldots,\Phi_{k+1}))&=-(-1)^{d(\Phi_1)}\omega_\text{c}(l_{k,0}(\Phi_1,\ldots,\Phi_k),\Phi_{k+1})\,.
\end{align}
The other couplings entering in $S_{0,1}$ always involve at least one open string and are given by
\begin{align}
 \mathcal{V}_{0,1}^{k,l+1} (\Phi^{\wedge k},\Psi^{\otimes l+1})= \frac{1}{k!(l+1)}\omega_\text{o}(\Psi,m_{k,l}(\Phi^{\wedge k};\Psi^{\otimes l}))\,,
\end{align}
where  
\begin{align}
m_{k,l}:(\mathcal{H}^\text{c})^{\wedge k}\otimes (\mathcal{H}^\text{o})^{\otimes l}\longrightarrow \mathcal{H}^\text{o}    
\end{align}
for $k,l\geqslant 0$, $k+l>0$ (that is, we set $m_{0,0}=0$ to avoid an open string tadpole) are degree $+1$ products which are graded-symmetric in their first $k$ entries and cyclic with respect to $\omega_\text{o}$ in their last $l$ entries, namely
\begin{align}
     \omega_\text{o}(\Psi_1,m_{k,l}(\Phi_1,\ldots,\Phi_{k};\Psi_2,\ldots,\Psi_{l+1}))&=-(-1)^{d(\Phi_1)+d(\Psi_1)[d(\Phi_1)+\ldots+d(\Phi_k)]}\times\nonumber\\
    &\hspace{0.4cm}\times\omega_\text{o}(m_{k,l}(\Phi_1,\ldots,\Phi_k;\Psi_1,\ldots,\Psi_l),\Psi_{l+1})\,.\label{opencyc}
\end{align}
For later convenience, let us also define the degree $+1$ products (for $k\geqslant 0$, $l\geqslant 1$)
\begin{align}
l_{k,l}:(\mathcal{H}^\text{c})^{\wedge k}\otimes (\mathcal{H}^\text{o})^{\otimes l}\longrightarrow \mathcal{H}^\text{c}    
\end{align}
by requiring that they satisfy
\begin{align}
    &\omega_\text{o}(\Psi,m_{k,l}(\Phi_1,\ldots,\Phi_k;\Psi_1,\ldots,\Psi_l)) =\nonumber\\
    &\hspace{1cm}= (-1)^{d(\Psi)[1+d(\Phi_1)+\ldots+d(\Phi_k)]+d(\Phi_1)}\omega_\text{c}(\Phi_1, l_{k-1,l+1}(\Phi_2,\ldots,\Phi_k;\Psi,\Psi_1,\ldots,\Psi_l))\,,\label{ocsym}
\end{align}
for all $\Phi_i\in\mathcal{H}^\text{c}$ and $\Psi,\Psi_j\in\mathcal{H}^\text{o}$. It is clear that such products $l_{k,l}$ are graded-symmetric in their first $k$ entries, cyclic in their last $l$ entries and cyclic with respect to $\omega_\text{c}$ in their first $k$ entries. 
Given this definition we can now compute the $b=1$ master equation \eqref{b1}. We start with
\begin{subequations}
\begin{align}
g_s(S_{0,0},S_{0,1})_\text{c} &= \sum_{k\geqslant 1}\sum_{m\geqslant 0}\frac{1}{k!}\frac{1}{m!}\omega_\text{c}(l_k(\Phi^{\wedge k}),l_{m,0}(\Phi^{\wedge m}))+\nonumber\\
    &\hspace{2cm}+\sum_{k,m,r\geqslant 1}\frac{1}{k!(m-1)!}\frac{1}{r}\omega_\text{c}(l_k(\Phi^{\wedge k}),l_{m-1,r}(\Phi^{\wedge (m-1)};\Psi^{\otimes r}))\\
    &= \sum_{k\geqslant 1}\sum_{m\geqslant 0}\frac{1}{k!}\frac{1}{m!}\Big[\frac{k}{k+m}\omega_\text{c}(l_k(\Phi^{\wedge k}),l_{m,0}(\Phi^{\wedge m}))+\frac{m}{k+m}\omega_\text{c}(l_{m,0}(\Phi^{\wedge m}),l_k(\Phi^{\wedge k}))\Big]+\nonumber\\
    &\hspace{2cm}+\sum_{k,m,r\geqslant 1}\frac{1}{k!(m-1)!}\frac{1}{r}\omega_\text{c}(l_k(\Phi^{\wedge k}),l_{m-1,r}(\Phi^{\wedge (m-1)};\Psi^{\otimes r}))\\
    &= -\sum_{r=0}^\infty\sum_{k=1}^{r+1}\frac{1}{(r+1)!}\omega_\text{c}(\Phi,[l_k, l_{r+1-k,0}](\Phi^{\wedge r}))+\nonumber\\
    &\hspace{1cm}-\sum_{r=1}^\infty \sum_{k=1}^r\sum_{s=1}^\infty\frac{1}{r!}\frac{1}{s}\omega_\text{o}(\Psi,m_{k,s-1}l_{r+1-k}(\Phi^{\wedge r};\Psi^{\otimes (s-1)}))\,.
\end{align}
\end{subequations}
Then we evaluate
\begin{subequations}
\label{eq:openclosed}
    \begin{align}
        g_s(S_{0,1},S_{0,1})_\text{o}&=\sum_{k,m\geqslant 0}\sum_{r,s\geqslant 1} \frac{1}{k!m!}\omega_\text{o}(m_{k,r-1}(\Phi^{\wedge k};\Psi^{\otimes (r-1)}),m_{m,s-1}(\Phi^{\wedge m};\Psi^{\otimes (s-1)}))\\
        &=-\sum_{r=0}^\infty \sum_{k=1}^{r+1}\frac{2}{(r+1)!} \omega_\text{c}(\Phi,l_{k-1,1}m_{r+1-k,0}(\Phi^{\wedge r}))+\nonumber\\
        &\hspace{0.4cm}-\sum_{r=0}^\infty\sum_{s=1}^\infty\sum_{k=0}^r\sum_{m=1}^{s+1}
        \frac{2}{r!}\frac{1}{s}\omega_\text{o}(\Psi,m_{k,m-1}m_{r-k,s+1-m}(\Phi^{\wedge r};\Psi^{\otimes (s-1)})))\,.
    \end{align}
\end{subequations}
Hence, the master equations $(S_{0,0},S_{0,0})_\text{c}=0$ and $2(S_{0,0},S_{0,1})_\text{c}+(S_{0,1},S_{0,1})_\text{o}=0$ can be solved by requiring the following Sphere-Disk Homotopy Algebra (SDHA)
\begin{subequations}
	\label{eq:SDHA}
\begin{align}
	0&=   \sum_{k=1}^{r-1} l_k l_{r-k}\,,\\
    0&= \sum_{k=1}^{r}[l_k, l_{r-k,0}]+\sum_{k=1}^{r-1}l_{k-1,1}m_{r-k,0}\,,\label{nonOCHA}\\
    0&= \sum_{k=1}^r m_{k,s-1}l_{r+1-k}+\sum_{k=0}^r\sum_{n=1}^{s}m_{k,n}m_{r-k,s-n}\,.\label{OCHA}
\end{align}
\end{subequations}
In particular, we recognize the third set of relations as those of Kajiura-Stasheff OCHA \cite{Kajiura:2004xu, Kajiura:2005sn}. The second set of relations are needed to solve the $b=1$ master equation, but notice that they do not have open string entries, therefore they will not be important for the local dynamics of the open strings in the action \eqref{action-def}.  However, as we will see, they will be fundamental for the correct construction of observables  and for the overall physics we will discuss. 
Notice that the products $m_{0,l}$ satisfy the $A_\infty$ relations of a pure open string field theory. As usual we will denote $l_1\equiv Q_\text{c}$ and $m_{0,1}\equiv Q_\text{o}$, the closed- and open-string BRST operators. Finally, note that the SDHA relations \eqref{eq:SDHA} can be equivalently recast in the form
\begin{subequations}
\begin{align}
0&=   \sum_{k=1}^{r-1} l_k l_{r-k}\,,\\
0&=   \sum_{n=1}^{s-1} m_{0,n} m_{0,s-n}\,,\\
0&= \sum_{k=1}^{r}[l_k, l_{r-k,s}]+\sum_{k=1}^{r}\sum_{n=0}^s l_{k-1,n+1}m_{r-k,s-n}\,,\label{eq:SDHAcompact}
\end{align}
\end{subequations}
which separates the $L_\infty$ and $A_\infty$ algebras satisfied by the sphere closed-string products $l_k$ and pure open-string products $m_{0,n}$ from the open-closed relation satisfied by the disk vertices with at least one closed string insertion.

\subsection{Open string field theory in a deformed closed string background}\label{sec:2.3}
Let us now analyze in more detail the OSFT action  $S_{\Phi_{\rm cl}}(\Psi)$ \eqref{action-def}. For this we fix a closed-string background by specifying a matter $\mathrm{CFT}_0$ and closed-string coupling constant $g_\mathrm{s}^{(0)}$, as well as an open-string background, given by some $\text{BCFT}_0$. We then write
\be
S_{\Phi_{\rm cl}}(\Psi)=Z_{\rm disk}^{(0)}+\Lambda_\text{c}(\Phi_{\rm cl}) -\frac{1}{g_s^{(0)}}\sum_{l=0}^\infty \frac1{l+1}\omega_\text{o}\left(\Psi,\tilde m_l(\Psi^{\ot l})\right)\,,
\ee
where 
\be
 \Lambda_\text{c}(\Phi_{\rm cl})= -\frac{1}{g_s^{(0)}}\sum_{k=0}^\infty \frac1{(k+1)!}\omega_\text{c}\left(\Phi_{\rm cl},l_{k,0}(\Phi_{\rm cl}^{\wedge k})\right)\label{lambdac}
\ee
is a non-dynamical constant and 
\be
\tilde m_l(\Psi_1\ot\cdots\ot\Psi_l)=\sum_{k=0}^\infty \frac1{k!} m_{k,l}(\Phi_{\rm cl}^{\wedge k};\Psi_1\ot\cdots\ot\Psi_l)
\ee
are cyclic open string products. Finally, we will choose a normalization in which we have
\begin{align}
	Z_\text{disk}^{(0)} = -\frac{1}{2\pi^2 g_s^{(0)}}\langle 0\| B_0\rangle\!\rangle\,,\label{Zdisk0}
\end{align}
where $\|B_0\rangle\!\rangle$ is the matter $\text{CFT}_0$ boundary state specifying our perturbative open-closed background $\text{BCFT}_0$.
Notice that there is a non vanishing zero product (tadpole) 
\be
\tilde m_0=\sum_{k=1}^\infty \frac1{k!} m_{k,0}(\Phi_{\rm cl}^{\wedge k})\in {\cal H}^\text{o}\,.
\ee
Since this action obeys the classical BV master equation \eqref{open-bv} its cyclic products necessarily form a (weak) $A_\infty$ algebra
\be
\sum_{l=0}^{r} \tilde m_l\tilde m_{r-l}=0,\quad r\geq 0\,.
\ee
This statement can be explicitly verified by using the OCHA relations \eqref{OCHA} and the fact that $\Phi_\mathrm{cl}$ satisfies the classical closed SFT equation of motion.
Therefore we have a consistent classical gauge invariant action with a tadpole. To find a stable vacuum we have to cancel the tadpole with a vacuum shift solution $\Psi_{\rm v}$ which is a solution to the weak $A_\infty$ equation of motion
\be
\sum_{l=0}^\infty \tilde m_l(\Psi_{\rm v}^{\ot l})=0\,.\label{veom}
\ee
Notice that, although not explicitly indicated, the vacuum shift solution $\Psi_{\rm v}$ generically depends on the closed string solution $\Phi_{\rm cl}$, that is $\Psi_\mathrm{v}\equiv \Psi_\mathrm{v}(\Phi_\mathrm{cl})$. If we are able to find such a solution we can write $\Psi=\Psi_{\rm v}+\psi$ and finally consider the theory for the fluctuations $\psi$. The resulting action will read
\be
S^{(*)}(\psi)=Z_{\rm disk}^{(0)}+ \Lambda_\text{c}(\Phi_{\rm cl})+\Lambda_\text{o}(\Phi_{\mathrm{cl}},\Psi_{\rm v})-\frac{1}{g_s^{(0)}}\sum_{l=1}^\infty \frac1{l+1}\omega_\text{o}\left(\psi,\hat m_l(\psi^{\ot l})\right)\,,\label{Sstar}
\ee
where a new constant term has been generated by the open string vacuum shift
\begin{subequations}
\begin{align}
\Lambda_\text{o}(\Phi_\mathrm{cl};\Psi_{\rm v})&=-\frac{1}{g_s^{(0)}}\sum_{l=0}^\infty \frac1{l+1}\omega_\text{o}\left(\Psi_{\rm v},\tilde m_l(\Psi_{\rm v}^{\ot l})\right)\\
&=-\frac{1}{g_s^{(0)}}\sum_{l=0}^\infty\sum_{k=0}^\infty \frac1{l+1}\frac{1}{k!}\omega_\text{o}\left(\Psi_{\rm v}, m_{k,l}(\Phi_\mathrm{cl}^{\wedge k};\Psi_{\rm v}^{\ot l})\right)
\,.
\end{align}
\label{lambdao}
\end{subequations}
Notice that the tadpole has disappeared thanks to the vacuum-shift equation \eqref{veom}
\be
\omega_\text{o}\left(\psi,\sum_{l=0}^\infty \tilde m_l(\Psi_{\rm v}^{\ot l})\right)=0
\ee
and that the new products $\hat m_l$ (whose structure can be readily written down by expanding and collecting terms with given power of $\psi$) now obey a standard $A_\infty$-algebra
\be
\sum_{l=1}^{r-1} \hat m_l\hat m_{r-l}=0\,,\quad r\geq 1\,.
\ee

\subsection{The cosmological constant and its physical interpretation}\label{sec:2.4}
In this paper we are mainly interested in understanding the total cosmological constant $\Lambda(\Phi_\mathrm{cl};\Psi_\mathrm{v})$ given by 
\begin{subequations}
	\label{Lambda}
\begin{align}
\Lambda(\Phi_{\mathrm{cl}};\Psi_{\rm v})&=\Lambda_\text{c}(\Phi_{\rm cl})+\Lambda_\text{o}(\Phi_{\mathrm{cl}};\Psi_{\rm v})\\
&= -\frac{1}{g_s^{(0)}}\sum_{k=0}^\infty \frac1{(k+1)!}\omega_\text{c}\left(\Phi_{\rm cl},l_{k,0}(\Phi_{\rm cl}^{\wedge k})\right)+\nonumber\\
&\hspace{2cm}-\frac{1}{g_s^{(0)}}\sum_{l=0}^\infty\sum_{k=0}^\infty \frac1{l+1}\frac{1}{k!}\omega_\text{o}\left(\Psi_{\rm v}, m_{k,l}(\Phi_\mathrm{cl}^{\wedge k};\Psi_{\rm v}^{\ot l})\right)\,.
\end{align}
\end{subequations}
 A necessary condition for this quantity to be physically meaningful is that it should remain invariant under the $L_\infty$ gauge transformations of $\Phi_\mathrm{cl}$, as well as under the weak $A_\infty$ gauge transformations of $\Psi_\mathrm{v}$. First of all, since $\Lambda_\text{o}(\Phi_{\mathrm{cl}};\Psi_{\rm v})$ is just the  weak-$A_\infty$ open string action evaluated on a classical open string solution, $\Lambda(\Phi_\mathrm{cl};\Psi_\mathrm{v})$ is obviously gauge invariant under an open string gauge transformation of $\Psi_{\rm v}$.
What is less obvious is what happens to $\Lambda(\Phi_\mathrm{cl};\Psi_\mathrm{v})$ if we perform a closed string gauge transformation of the closed string classical solution $\Psi_{\rm cl}$. We will now show that $\Lambda(\Phi_\mathrm{cl};\Psi_\mathrm{v})$ is gauge invariant under closed string gauge transformations.

\subsubsection{Gauge invariance of $\Lambda(\Phi_\mathrm{cl};\Psi_\mathrm{v})$}

Let us consider the gauge variation 
\begin{align}
	\delta_\Omega \Phi_{\mathrm{cl}} = Q_c \Omega + l_2(\Omega,\Phi_\mathrm{cl})+ \frac{1}{2!} l_3(\Omega,\Phi_\mathrm{cl},\Phi_\mathrm{cl})+\ldots = \sum_{n=1}^\infty \frac{1}{(n-1)!}l_n(\Omega,\Phi_\mathrm{cl}^{\wedge (n-1)})\label{deltaphi}
	\end{align}
of the closed SFT classical solution $\Phi_\mathrm{cl}$, where the degree-odd closed string field $\Omega$ is the corresponding gauge parameter. Considering first the variation of $\Lambda_\mathrm{o}(\Phi_\mathrm{cl};\Psi_\mathrm{v})$, we have
\begin{align}
-\delta_\Omega\Lambda_\mathrm{o}(\Phi_\mathrm{cl};\Psi_\mathrm{v}) &= \frac{1}{g_s^{(0)}}\sum_{l=0}^\infty\sum_{k=1}^\infty \frac1{l+1}\frac{1}{(k-1)!}\omega_\text{o}\left(\Psi_{\rm v}, m_{k,l}(\delta_\Omega\Phi_\mathrm{cl},\Phi_\mathrm{cl}^{\wedge (k-1)};\Psi_{\rm v}^{\ot l})\right)+\nonumber\\
&\hspace{4cm}+\frac{1}{g_s^{(0)}}\sum_{l=0}^\infty\sum_{k=0}^\infty \frac{1}{k!}\omega_\text{o}\left(\delta_\Omega\Psi_{\rm v}, m_{k,l}(\Phi_\mathrm{cl}^{\wedge k};\Psi_{\rm v}^{\ot l})\right)\,,
\end{align}
where the second term clearly vanishes due to the weak $A_\infty$ equation of motion \eqref{veom} satisfied by $\Psi_\mathrm{v}$. Isolating the closed-string variation $\delta_{\Omega}\Phi_\mathrm{cl}$ in the bra of the symplectic form and substituting, we therefore obtain
\begin{subequations}
\begin{align}
-\delta_\Omega\Lambda_\mathrm{o}(\Phi_\mathrm{cl};\Psi_\mathrm{v}) &= \frac{1}{g_s^{(0)}}\sum_{l=0}^\infty\sum_{k=1}^\infty \frac1{l+1}\frac{1}{(k-1)!}\omega_\text{c}\left( \delta_\Omega\Phi_\mathrm{cl},l_{k-1,l+1}(\Phi_\mathrm{cl}^{\wedge (k-1)};\Psi_{\rm v}^{\ot l+1})\right)\\
&= \frac{1}{g_s^{(0)}}\sum_{l=0}^\infty\sum_{k=1}^\infty\sum_{n=1}^\infty \frac1{l+1}\frac{1}{(k+n-2)!}\times\nonumber\\
&\hspace{2cm}\times
\omega_\text{c}\left( \Omega,l_n l_{k-1,l+1}(\Phi_\mathrm{cl}^{\wedge (k+n-2)};\Psi_{\rm v}^{\ot l+1})\right)\\
&= \frac{1}{g_s^{(0)}}\sum_{l=0}^\infty\sum_{k=1}^\infty\sum_{n=1}^\infty \frac1{l+1}\frac{1}{(k+n-2)!}\times\nonumber\\
&\hspace{2cm}\times
\omega_\text{c}\left( \Omega,[l_n, l_{k-1,l+1}](\Phi_\mathrm{cl}^{\wedge (k+n-2)};\Psi_{\rm v}^{\ot l+1})\right)\,,
\end{align}
\end{subequations}
where we have used the closed SFT equation of motion to obtain the last line. At the same time, we can vary the closed-string part $\Lambda_\mathrm{c}(\Phi_\mathrm{cl})$ of the cosmological constant to obtain
\begin{subequations}
	\begin{align}
	-\delta_\Omega\Lambda_\mathrm{c}(\Phi_\mathrm{cl}) &=\frac{1}{g_s^{(0)}}\sum_{k=1}^\infty \frac1{(k-1)!}\omega_\text{c}\left(\delta_\Omega\Phi_{\rm cl},l_{k-1,0}(\Phi_{\rm cl}^{\wedge k-1})\right)\\
	&=\frac{1}{g_s^{(0)}}\sum_{k=1}^\infty \sum_{n=1}^\infty \frac{1}{(k+n-2)!}\omega_\text{c}\left( \Omega,[l_n, l_{k-1,0}](\Phi_\mathrm{cl}^{\wedge k+n-2})\right)\,,
	\end{align}
\end{subequations}
again using the closed SFT equation of motion to obtain the last equality. Altogether we can therefore write
\begin{align}
-\delta_\Omega\Lambda (\Phi_\mathrm{cl};\Psi_\mathrm{v}) &= \frac{1}{g_s^{(0)}}\sum_{r=1}^\infty\sum_{k=1}^{r}\sum_{s=0}^\infty \frac{f(s)}{(r-1)!}
\omega_\text{c}\left( \Omega,[l_{k}, l_{r-k,s}](\Phi_\mathrm{cl}^{\wedge r-1};\Psi_{\rm v}^{\ot s})\right)\,,
\end{align}
where $f(s)=1/s$ for $s\geq 1$ and $f(0)=1$. Using now the SDHA relations \eqref{eq:SDHAcompact}, we can rewrite this as\footnote{In general, using cyclicity of $l_{k,n}$ in their open-string arguments, we would write
\begin{align}
&l_{k-1,n+1}m_{l,m}(\Phi_1,\ldots,\Phi_{k+l-1};\Psi_1,\ldots,\Psi_{m+n})=\nonumber\\
&\hspace{0.2cm} = \sum_{\sigma}\sum_{j=1}^{m+n}  l_{k-1,n+1}(\Phi_{\sigma(1)},\ldots,\Phi_{\sigma(k-1)};
m_{l,m}(\Phi_{\sigma(k)},\ldots,\Phi_{\sigma(k+l-1)};\Psi_{j},\ldots,\Psi_{j+m-1}),\ldots,\Psi_{j+m+n})\,,
\end{align}
where $\sigma$ runs over all possible ways of picking unordered $k-1$ elements out of a pool of $k+l-1$. Upon setting $\Phi_1=\ldots=\Phi_{k+k-1}$ and $\Psi_1=\ldots=\Psi_{m+n}$, the sums produce the required factor of $(m+n)(k+l-1)!/[(k-1)!l!]$.
}
\begin{align}
-\delta_\Omega\Lambda (\Phi_\mathrm{cl};\Psi_\mathrm{v}) &= -\frac{1}{g_s^{(0)}}\sum_{k=1}^{\infty}\sum_{l=0}^\infty\sum_{m=0}^\infty \sum_{n=0}^\infty \frac{1}{(k-1)!}\frac{1}{l!}\times\nonumber\\
&\hspace{2cm}\times
\omega_\text{c}\left( \Omega, l_{k-1,n+1}(\Phi_\mathrm{cl}^{\wedge k-1};m_{l,m}(\Phi_\mathrm{cl}^{\wedge l};\Psi_{\rm v}^{\ot m}),\Psi_{\rm v}^{\ot n})\right)\,,
\end{align}
which vanishes as a consequence of the weak $A_\infty$ equation of motion \eqref{veom} satisfied by $\Psi_\mathrm{v}$.
This shows that the total cosmological constant $\Lambda(\Phi_\mathrm{cl};\Psi_\mathrm{v})$ is indeed invariant under the closed SFT gauge transformation of $\Phi_\mathrm{cl}$.

\subsubsection{Physical meaning of $\Lambda(\Phi_\mathrm{cl};\Psi_\mathrm{v})$}
The theory for the fluctuations  \eqref{Sstar} should  be equivalent by field redefinition to an open string field theory action for a new D-brane system $\mathrm{BCFT}_{*}$ with matter boundary state $\|B_\ast\rangle\!\rangle$, described by $\Psi_{\rm v}$, in a new closed string background ($\mathrm{CFT}^{(*)}$, $g_s^{(\ast)}$) described by $\Phi_{\rm cl}$. Therefore, as a first step, we should  be able to rewrite the action \eqref{Sstar} in the form
\begin{align}
	S^{(*)}(\psi)=Z_\mathrm{disk}^{(*)}-\frac{1}{g_s^{(\ast)}}\sum_{l=1}^\infty \frac1{l+1}\omega_\text{o}\left(\psi, m_l^{(\ast)}(\psi^{\ot l})\right)\,,\label{pop}
	\end{align}
where $g_s^{(\ast)}$ is the closed-string coupling constant in the new background and 
\begin{align}
	Z_\mathrm{disk}^{(*)} = -\frac{1}{2\pi^2 g_s^{(\ast)}} \langle 0\|B_\ast\rangle\!\rangle
	\end{align}
is the new disk partition function.  This identification is meaningful because $Z_\mathrm{disk}^{(*)}$  in \eqref{pop} is invariant under field redefinitions  of $\psi$ which don't include a constant shift (which, in general, would recreate a tadpole).  Then, from \eqref{Sstar}, we can read off 
\begin{align}
Z_\mathrm{disk}^{(\ast)}= Z_{\rm disk}^{(0)}+  \Lambda(\Phi_{\mathrm{cl}};\Psi_{\rm v})\,.\label{Zstar}
\end{align}
In addition, we should also identify
\begin{align}
m_l^{(*)} = \frac{g_s^{(*)}}{g_s^{(0)}}\,\hat{m}_l\,,
\end{align}
where we note that rescaling the products by a constant factor which is independent of $l$ does not spoil the $A_\infty$ algebra they satisfy. Notice however that these products still act on the original degrees of freedom of the initial open-closed background and only after a field redefinition (which is an $A_\infty$ morphism) they will give rise to $A_\infty$ products on the new background.

As it is apparent from \eqref{Zstar}, $\Lambda(\Phi_\mathrm{cl};\Psi_\mathrm{v})$ should be thus interpreted as the shift in the worldsheet disk partition function $Z_\mathrm{disk}$ between the perturbative background given by the data $(\mathrm{CFT}_0, g_s^{(0)},\|B_0\rangle\!\rangle)$ and the new background $(\mathrm{CFT}_\ast, g_s^{(\ast)},\|B_\ast\rangle\!\rangle)$ corresponding to $\Phi_\mathrm{cl}$ and $\Psi_\mathrm{v}$. 
In general, this will encompass both a shift in the $g$-function of the matter boundary state, as well as a possible shift in the value of the string coupling constant $g_s$, as the disk partition function depends on both of these parameters (see \eqref{Zdisk0}). In the remainder of this paper we will present a number of arguments supporting this conjecture.

First, let us consider the trivial case when $\Phi_{\rm cl}=0$. Then we can take $\Psi_{\rm v}$ to be a classical solution of a standard $A_\infty$ open SFT (without a tadpole) which will describe another D-brane system in the same closed string background. The cosmological constant $\Lambda(\Phi_\mathrm{cl}=0,\Psi_\mathrm{v})=\Lambda_\mathrm{o}(\Phi_\mathrm{cl}=0,\Psi_\mathrm{v})$ is then equal to the open SFT action evaluated on the classical solution $\Psi_\mathrm{v}$. Hence, by the Sen's conjecture, the total cosmological constant can be expressed as (normalizing the identity matter correlator to be equal to the matter $g$-function and setting $\alpha^\prime =1$)
\begin{align}
\Lambda(\Phi_\mathrm{cl}=0,\Psi_\mathrm{v}) = \frac{1}{2\pi^2 g_s^{(0)}}\Big(\langle 0\| B_0\rangle\!\rangle-\langle 0\| B_\ast \rangle\!\rangle\Big)\,,
\end{align}
which indeed reproduces the required shift in $Z_\mathrm{disk}$ with $g_s^{(\ast)}=g_s^{(0)}$ (which makes sense because setting $\Phi_\mathrm{cl}=0$  guarantees that we are not changing the string coupling constant). In particular, for the tachyon vacuum solution, we will find $Z_\mathrm{disk}^{(\ast)}=0$. We conclude that the OSFT Sen's conjecture is a special case of our more general conjecture \eqref{Zstar}.

%
%
\section{Perturbative background shifts}\label{sec:3}
In the rest of the paper, we will consider non-trivial solutions $\Phi_\mathrm{cl}\equiv \Phi_\mu$ (where $\mu$ is some continuous parameter) which correspond to marginal deformations of the perturbative closed string background. More concretely, in this section we will review the standard order-by-order construction of $\Phi_\mu$ in Siegel gauge and, when it will exist, we will construct the associated perturbative vacuum shift solution $\Psi_\mathrm{v}\equiv \Psi_\mu$, which adapts the initial D-brane system $\|B_0\rangle\!\rangle$ to the new closed string background described by $\Phi_\mu$. Using $\mathrm{SL}(2,\mathbb{C})$ maps, we will also provide explicit expressions for the open-closed vertices which are needed for the computation of the cosmological constant at linear and quadratic order in the closed string modulus $\mu$.

\subsection{Closed string solution}
\label{sec:3.1}

We start considering  a closed string classical solution generated by an element of the ${\rm gh}=2$ closed string cohomology $V$, namely  
\be
\Phi_{\rm cl}\equiv \Phi_\mu=\mu V+\sum_{k=2}^\infty \mu^k\,\Phi_k\,.
\ee
The components $\Phi_k$ of the solution at order $\mu^k$ are then fixed by solving the closed SFT equation of motion 
\be
\sum_{k=1}^\infty\,\frac1{k!} l_k(\Phi^{\wedge k})=0,
\ee
order by order in $\mu$. For the sake of definiteness, we will construct the solution in Siegel gauge $b_0^+=0$ so that
\begin{subequations}
\begin{align}
\Phi_2&=-\frac12\frac{b_0^+}{L_0^+}\bar{P}_0^+ l_2(V,V)\,,\\
\Phi_3&= \frac{b_0^+}{L_0^+}\bar{P}_0^+\left(\frac14l_2\left(V,\frac{b_0^+}{L_0^+}\bar{P}_0^+l_2(V,V)\right)-\frac16l_3(V,V,V)\right)\,.\\
&\hspace{0.2cm}\vdots\nonumber
\end{align}
\end{subequations}
This is a solution provided the propagators always act outside of the kernel of $L_0^+$ which, calling $P_0^+$ the projector on the kernel of $L_0^+$, means that we should ensure
\begin{subequations}
\begin{align}
0&=P^+_0l_2(V,V)\,,\\
0&=P^+_0\left(\frac12l_2\left(V,\frac{b_0^+}{L_0^+}\bar{P}_0^+ l_2(V,V)\right)-\frac13l_3(V,V,V)\right)\,.\\
&\hspace{0.2cm}\vdots\nonumber
\end{align}
\end{subequations}
This is a statement about the vanishing of sphere amplitudes involving an arbitrary number of $V$'s and any other state in the kernel of $L_0^+$, which in turn implies that the effective potential for $V$ should vanish \cite{Erbin:2020eyc}. With the marginal closed string solution at our disposal, we can construct the closed string part of the cosmological constant \eqref{lambdac}, which to second order in the modulus $\mu$, is given by
\be
-g_s^{(0)}\Lambda_\text{c}(\mu)=\mu\, \omega_\text{c}(V,l_{0,0})+\frac12\mu^2\,\left(\omega_\text{c}\!\left(l_{0,0}\,,\frac{b_0^+}{L_0^+}\bar{P}_0^+l_2(V,V)\right)+\omega_\text{c}\!\left(V,l_{1,0}(V)\right)\right)+\mathcal{O}(\mu^3)\,.
\ee

\subsection{Open string vacuum-shift solution}
\label{sec:3.2}

To construct the open string part of the cosmological constant \eqref{lambdao} we need to provide a solution $\Psi_\mathrm{v}$ to the weak $A_\infty$ equation of motion \eqref{veom}. We will attempt to find such $\Psi_\mathrm{v}\equiv \Psi_\mu$ perturbatively in $\mu$. If successful, this means that the corresponding new D-brane system is very close to the old one, meaning that this particular D-brane system can adapt to the changes in the closed string background dictated by $\Phi_\mathrm{cl}$. If this fails, one can assume that it is still possible to find a non-perturbative solution $\Psi_\mathrm{v}$ to \eqref{veom} which, however, would generally describe a D-brane system far from the initial one.

To calculate $\Psi_\mu$, we have to solve the tadpole-sourced equation of motion
\begin{align}
\tilde m_0+\tilde m_1(\Psi)+\tilde m_2(\Psi,\Psi)+\ldots=0\,.
\end{align}
We search for a perturbative vacuum shift 
\be
\Psi_{\rm v}\equiv\Psi_\mu=\sum_{l=1}^\infty \mu^l\,\Psi_l\,.
\ee
Expanding in $\mu$ the products $\tilde m_k$ using the closed string solution $\Phi_\mu$,  we get the recursive equations
\begin{subequations}
\begin{align}
Q\Psi_1&=-m_{1,0}(V)\label{shift-1}\,,\\
Q\Psi_2&=-\frac12\left[m_{2,0}(V,V)-m_{1,0}\left(\frac{b_0^+}{L_0^+}\bar{P}_0^+ l_2(V,V)\right)\right]-m_{1,1}(V,\Psi_1)-m_{0,2}(\Psi_1,\Psi_1)\,.\\
&\hspace{0.2cm}\vdots\nonumber
\end{align}
\end{subequations}
From here it is apparent that, even if the closed string solution $\Phi_\mu$ is non-obstructed this does not mean that a perturbative open-string vacuum shift solution $\Psi_\mu$ necessarily exists. This is in fact completely analogous to what we have already discussed in \cite{Maccaferri:2021ksp,Erbin:2020eyc}. For example, looking at \eqref{shift-1} we see that $\Psi_1$ can only exist if $m_{1,0}(V)$ is BRST exact as an open string state (although $V$, as a bulk field, should be in the closed string BRST cohomology).  If this is the case, then we can write
\begin{subequations}
\begin{align}
\Psi_1&=-\frac{b_0}{L_0}\bar{P}_0 m_{1,0}(V)\,,\\
\Psi_2&=\frac{b_0}{L_0}\bar{P}_0 \bigg\{-\frac12\left[m_{2,0}(V,V)-m_{1,0}\left(\frac{b_0^+}{L_0^+}l_2(V,V)\right)\right]+\nonumber\\
&\hspace{7cm}-m_{1,1}(V,\Psi_1)-m_{0,2}(\Psi_1,\Psi_1)\bigg\}\,,\\
&\hspace{0.2cm}\vdots\nonumber
\end{align}
\end{subequations}
which is a solution provided the corresponding obstructions vanish
\begin{subequations}
\begin{align}
0&=P_0m_{1,0}(V)\,,\\
0&=P_0\bigg\{-\frac12\left[m_{2,0}(V,V)-m_{1,0}\left(\frac{b_0^+}{L_0^+}\bar{P}_0^+ l_2(V,V)\right)\right]+\nonumber\\
&\hspace{7cm}-m_{1,1}(V,\Psi_1)-m_{0,2}(\Psi_1,\Psi_1)\bigg\}\,.\\
&\hspace{0.2cm}\vdots\nonumber
\end{align}
\end{subequations}
As explained in \cite{Maccaferri:2021ksp} the vanishing of these obstructions is a statement about the vanishing of amplitudes between a single open string in the kernel of $L_0$  and any number of deforming closed strings $V$. This   means in turn that the effective action for  open strings in $\mathrm{ker}\,L_0$ has no tadpole. 
The absence of this `massless' tadpole means that the initial D-brane system can continuously adapt to the background deformation. Notice that these amplitudes, by construction, use both open and closed string propagators. This is a major difference with respect to \cite{Maccaferri:2021ksp} where the moduli space was covered with only open-string propagators.

Now we can write down the leading-order contribution to the open string part of the cosmological constant which turns out to be
\begin{align}
-g_s^{(0)}\Lambda_\text{o}(\mu)=\frac12 \mu^2 \omega_\text{o}\left(m_{1,0}(V),\frac{b_0}{L_0}\bar{P}_0 m_{1,0}(V)\right)+\mathcal{O}(\mu^3).
\end{align}
Notice that, as in the case of Witten theory deformed by the Ellwood invariant \cite{Maccaferri:2021ksp}, $\Lambda_\text{o}$ starts at $\mathcal{O}(\mu^2)$ and the $\mathcal{O}(\mu)$ contribution is completely stored in $\Lambda_\text{c}$, which is invisible in Witten theory.

\subsection{Total cosmological constant}
\label{sec:3.3}

If we collect all  contributions  in \eqref{Lambda} we find the total cosmological constant
\begin{align}
-g_s^{(0)}\Lambda(\mu)&= \mu\, \omega_\text{c}(V,l_{0,0})+\frac12\mu^2\,\bigg[\omega_\text{c}\!\left(l_{0,0}\,,\frac{b_0^+}{L_0^+}\bar{P}_0^+ l_2(V,V)\right)+\omega_\text{c}\!\left(V,l_{1,0}(V)\right)+\nonumber\\
&\hspace{5.4cm}+\omega_\text{o}\!\left(m_{1,0}(V),\frac{b_0}{L_0}\bar{P}_0m_{1,0}(V)\right)\bigg] + \mathcal{O}(\mu^3)\,.\label{cosmo-pert}
\end{align}
One could continue to order $\mu^n$  to verify that this is the amplitude of $n$ deforming closed strings $V$ on the disk. In particular, starting at $\mathcal{O}(\mu^2)$, we see that the amplitude gets contribution from both closed and open constituents which, alone, would be inconsistent. To substantiate this fact, consider the amplitude of two physical  closed strings $V_1$ and $V_2$ that emerges from \eqref{cosmo-pert}  
\begin{align}
{\cal A}(V_1,V_2)&=\omega_\text{c}\!\left(l_{0,0}\,,\frac{b_0^+}{L_0^+}\bar{P}_0^+ l_2(V_1,V_2)\right)+\omega_\text{c}\!\left(V_1,l_{1,0}(V_2)\right)+\nonumber\\
&\hspace{7cm}+\omega_\text{o}\!\left(m_{1,0}(V_1),\frac{b_0}{L_0}\bar{P}_0m_{1,0}(V_2)\right).\label{ampl}
\end{align}
This is a correct amplitude because we can choose $V_2=Q_\text{c}\lambda$ and verify that it vanishes (up to possible terms at the boundary of moduli space) thanks to the nontrivial SDHA  homotopy relation
\begin{align}
[Q_\text{c},l_{1,0}]+l_2 l_{0,0}+&l_{0,1}m_{1,0}=0\,,\label{homrell}
\end{align}
as it is easy to explicitly check.

\subsection{Explicit open-closed vertices}
\label{sec:3.4}
Looking at the amplitude \eqref{ampl} we recognize that the part with the closed string propagator contains the region of closed string degeneration where the two closed strings collide while the part with the open string propagator contains the region of open string degeneration, where one of the closed strings hits the world-sheet boundary. The piece in the middle containing the $l_{1,0}$ vertex is responsible for the integration in the interior of the moduli space and does not contain IR divergences. Interestingly, it turns out that we can construct $l_2, l_{0,0}$ and $m_{1,0}$ in such a way that the homotopy relation \eqref{homrell} holds with $l_{1,0}=0$, namely
\begin{align}
l_{1,0}=0\quad\to\quad 
l_2 l_{0,0}+l_{0,1}m_{1,0}=0\,.\label{pippo}
\end{align}
In appendix \ref{app:A}, we give an explicit construction of these vertices using $\mathrm{SL}(2,\mathbb{C})$ maps which we summarize here for convenience:
\begin{subequations}
\begin{align}
l_2(\Phi_1,\Phi_2)&=b_0^-\delta(L_0^-) f_1\circ \Phi_1(0,\bar 0)\, f_2\circ \Phi_1(0,\bar 0) \ket0_{\mathrm{SL}(2,\mathbb{C})}\,, \\[2mm]
l_{0,0}&=\frac{1}{(2\pi i )^2}\lc^{-L_0^+}\| B_0\rangle\!\rangle\,,\\
(-1)^{d(\Phi)}m_{1,0}(\Phi)&=\frac{1}{2\pi i }\Big[\widetilde{m\circ \Phi(0,\bar 0)}\Big]\,\ket0_{\mathrm{SL}(2,\mathbb{R})}\,,
\end{align}
\end{subequations}
where $\widetilde{\cdots}$ means that doubling trick is understood and we have defined $\mathrm{SL}(2,\mathbb{C})$ functions
\begin{subequations}
\begin{align}
f_1(w)&=\frac1{\lc}\frac{w-\lc}{3\lc+w}=-f_2(w)\,,\\
m(w)&=\frac i\lo\frac{1+\frac{w}{\beta_2}}{1+\frac{w}{\beta_1}}\,.
\end{align}
\end{subequations}
To satisfy the homotopy relation \eqref{pippo} we have to relate the open and closed string parameters as (see appendix \ref{app:A})
\begin{subequations}
\begin{align}
\lo&=\frac{3\lc^2+1}{3\lc^2-1}\,,\label{openvsclosed}\\
\beta_2&=\frac{3\lc^2-1}{\lc^2+1}\,\lc\,,\\
\beta_1&=\frac{3\lc^2+1}{\lc^2-1}\,\lc\,.
\end{align}
\end{subequations}
This construction (which is obviously not unique) contains  a free parameter $\lc>1$  which selects how much moduli space of the amplitude \eqref{ampl} (with $l_{1,0}=0$) is covered by the first term (involving closed string propagation) or by the third term (involving open string propagation), as we are going to see explicitly in the next subsection.

\subsection{Explicit computation of the cosmological constant up to second order for a matter deformation}\label{sec:3.5}
We will now consider the case where the physical deforming closed string state $V$ is of the standard form 
\begin{equation}
V(z,\bar z)=c\bar c \V_{1,1}(z,\bar z)\,,
\end{equation}
where $\V_{1,1}$ is an $h=(1,1)$ matter primary. An analogous computation has been done in \cite{Maccaferri:2021ksp} in the context of Witten theory with Ellwood Invariant. Here we will appreciate how the present setting makes everything more rigorous but at the same time consistent with the conclusions reached in \cite{Maccaferri:2021ksp}. We use the Schwinger representation of the closed string propagator 
\be
\frac{b_0^+}{L_0^+}=\frac{b_0^+}{L_0^++\ec}{\Bigg |}_{\ec\to 0}=b_0^+\int_0^1\frac{dt}{t} \,t^{\ec}\,t^{L_0^+}{\Bigg |}_{\ec\to 0}\,,
\ee
where the closed string regulator $\ec$ is used to tame singularities at closed string degeneration $t\to0^+$ \cite{Maccaferri:2021ksp, Larocca:2017pbo, Sen:2019jpm}.
The closed string channel contribution is thus
\begin{subequations}
\begin{align}
&\wc\left(l_{0,0}\,,\frac{b_0^+}{L_0^+}\bar{P}_0^+l_2(V,V)\right)=\nonumber\\
&\hspace{2cm}=-\frac{1}{4\pi^2}\int_0^1 \frac{dt}{t}\, t^{\ec} \langle\!\langle B_0\| \left(\frac t\lc\right)^{L_0^+}b_0^+\,f_1\circ V(0,\bar 0)f_2\circ V(0,\bar 0)\rangle{\Bigg |}_{\ec\to 0}\\
&\hspace{2cm}=-\frac{1}{4\pi^2}\int_0^{\frac{1}{\lc}}\frac{dx}{x} x^{\ec} \aver{b_0^+ \, f_{1x}\circ V(0,\bar 0)f_{2x}\circ V(0,\bar 0)}_{\rm disk}{\Bigg |}_{\ec\to 0}\,,
\end{align}
\end{subequations}
where 
\be
f_{1x}(w)=\frac{x}{\lc}\frac{w-\lc}{3\lc+w}=-f_{2x}(w)\,.
\ee
Now, using that $V$ is a $(0,0)$ primary we can explicitly write
\begin{subequations}
\begin{align}
&\wc\left(l_{0,0}\,,\frac{b_0^+}{L_0^+}\bar{P}_0^+\,l_2(V,V)\right)=\nonumber\\
&\hspace{2cm}=-\frac{1}{4\pi^2}\int_0^{\frac1\lc}\frac{dx}{x}\,x^{\ec}\aver{b_0^+\,V\left(-\frac x{3\lc},-\overline{\frac x{3\lc}}\right)\,V\left(\frac x{3\lc},\overline{\frac x{3\lc}}\right)}_{\rm disk}{\Bigg |}_{\ec\to 0}\\
&\hspace{2cm}=-\frac{1}{4\pi^2}\int_0^{\frac{1}{3\lc^2}}\frac{dy}{y}\,y^{\ec}\aver{b_0^+ V(-y,-\bar y)V(y,\bar y)}_{\rm disk}{\Bigg |}_{\ec\to 0}\,.\label{closedprop}
\end{align}
\end{subequations}
The other contribution involving open string propagation reads instead (after using the analogous representation for the open string propagator)
\begin{align}
\wo\left(m_{1,0}(V),\frac{b_0}{L_0}\bar{P}_0m_{1,0}(V)\right)=\frac{1}{4\pi^2}\int_0^1\frac{dt}t\,t^{\eo}\,\aver{I_o\circ m\circ V(0,\bar 0)\,b_0\,t^{L_0}\,m\circ V(0,\bar 0)}_{\rm UHP}{\Bigg |}_{\eo\to 0},
\end{align}
where $\eo$ regulates the open string degeneration region $t\to 0$ and $I_o(z)=-1/z$. Again, using that $V$ is a $(0,0)$ primary this can be further simplified to 
\begin{subequations}
\begin{align}
\wo\left(m_{1,0}(V),\frac{b_0}{L_0}\bar{P}_0m_{1,0}(V)\right)&=\frac{1}{4\pi^2}\int_0^1\frac{dt}t\,t^{\eo}\,\aver{ V(i\lo,\bar {i\lo})\,b_0\,t^{L_0}\, V(i/\lo,\bar {i/\lo})}_{\rm UHP}{\Bigg |}_{\eo\to 0}\\
&=\frac{1}{4\pi^2}\int_0^1\frac{dt}t\,t^{\eo}\,\aver{ V(i,\bar {i})\,b_0\, V(it/\lo^2,\bar {it/\lo^2})}_{\rm UHP}{\Bigg |}_{\eo\to 0}\\
&=\frac{1}{4\pi^2}\int_0^{\frac1{\lo^2}}\frac{ds}{s}\,s^{\eo}\aver{V(i,\bar i) \,b_0\,V(is,\bar{is}}_{\rm UHP}{\Bigg |}_{\eo\to 0}\,,\label{openprop}
\end{align}
\end{subequations}
where in going from the first to second line we have used the scale invariance of the UHP. 
To connect the two contributions from closed \eqref{closedprop} and open \eqref{openprop} string propagation, we now recall the result proved in \cite{Maccaferri:2021ksp} (appendix A) 
\begin{align}
\frac{dy}y\aver{b_0^+ V(-y,-\bar y)V(y,\bar y)}_{\rm disk}=\frac{ds}{s}\aver{V(i,\bar i) \,b_0\,V(is,\bar{is})}_{\rm UHP}\,,
\end{align}
when we change integration variable
\be
y=\frac{1-\sqrt s}{1+\sqrt s}\quad\leftrightarrow\quad s=\left(\frac{1-y}{1+y}\right)^2.
\ee
Then we can write everything using the open string variable $s$ obtaining
\begin{subequations}
\begin{align}
&-\wo\left(m_{1,0}(V),\frac{b_0}{L_0}\bar{P}_0m_{1,0}(V)\right)-\wc\left(l_{0,0}\,,\frac{b_0^+}{L_0^+}\bar{P}_0^+l_2(V,V)\right)=\nonumber\\
&\hspace{2cm}=-\frac{1}{4\pi^2}\int_0^{\frac1{\lo^2}}\frac{ds}{s}\,s^{\eo}\aver{V(i,\bar i) \,b_0\,V(is,\bar{is}}_{\rm UHP}{\Bigg |}_{\eo\to 0}+\nonumber\\
&\hspace{4cm}-\frac{1}{4\pi^2}
\int_{\left(\frac{3\lc^2-1}{3\lc^2+1}\right)^2}^1\frac{ds}{s} (y(s))^{\ec}\aver{V(i,\bar i) \,b_0\,V(is,\bar{is}}_{\rm UHP}{\Bigg |}_{\ec\to 0}\\
&\hspace{2cm}=-\frac{1}{4\pi^2}\int_0^{\frac1{\lo^2}}\frac{ds}{s}\,s^{\eo}\aver{V(i,\bar i) \,b_0\,V(is,\bar{is}}_{\rm UHP}{\Bigg |}_{\eo\to 0}+\nonumber\\
&\hspace{4cm}-\frac{1}{4\pi^2}
\int_{\frac1{\lo^2}}^1\frac{ds}{s} (y(s))^{\ec}\aver{V(i,\bar i) \,b_0\,V(is,\bar{is}}_{\rm UHP}{\Bigg |}_{\ec\to 0}\\
&\hspace{2cm}=\frac{1}{\pi^2}\left(\int_0^{1/{\lo}^{2}} ds \,s^{\eo}+\int_{1/{\lo}^{2}}^1 ds \,y(s)^{\ec}\right)(1-s^2)\times\nonumber\\
&\hspace{6cm}\times\aver{\V_{1,1}(i,\bar{i})\V_{1,1}(is,\bar{is})}_{\rm UHP}{\Bigg |}_{\hspace{-1.7mm}\footnotesize\begin{array}{c}\ec\to 0\\ \eo\to 0\end{array}},\label{amplfin}
\end{align}
\end{subequations}
where we have used the relation between the open and closed stubs \eqref{openvsclosed} and we have evaluated the universal ghost correlator. Notice that this is independent of $\lo$ as can be readily seen by differentiating \eqref{amplfin} with respect to $\lo$ and taking the limits $\ec\to 0$ and $\eo\to 0$ (which are safe to take provided that we consider a fixed value $\lo>1$).
In particular, notice that for $\lo\to1^+$ (which means $\lc\to \infty$) the integration region of the second term in \eqref{amplfin} shrinks to zero and the whole moduli space is covered by the open string exchange, which is exactly what happens in Witten theory \cite{Maccaferri:2021ksp}. This limit is however delicate because of the regularization at $s\to 1$ which is not naturally included in the first term. Thus in the singular Witten theory we have to `remember'  to regulate the divergence at $s\to 1$ with the regulator $$y(s)^{\ec}=\left(\frac{1-\sqrt s}{1+\sqrt s}\right)^{\ec}\,,$$ as discussed in \cite{Maccaferri:2021ksp}.

\section{Example: Moving in the moduli space of a Narain compactification}\label{sec:4}

In this section we will evaluate the cosmological constant (up to quadratic order in the deformation parameter) for the special case of marginally deforming a $d$ dimensional Narain compactification given by the CFT of $d$ free bosons $X^1,\ldots, X^d$ on a lattice whose moduli are encoded in the tensor $E_{ij} = g_{ij}+B_{ij}$, where we also impose some conformally consistent boundary conditions. We will check that the sum of the disk partition function and the SFT cosmological constant factorizes into a universal part, which is independent of the boundary state $\|B(E)\rangle\!\rangle$ (and which, according to our conjecture, should indicate a change in the string coupling constant) and a part which precisely reproduces the change in the $g$-function $\langle 0\|B(E)\rangle\!\rangle$ of the corresponding boundary state (which can be calculated independently from the known form of $\|B(E)\rangle\!\rangle$). For the sake of concreteness, here we will explicitly work out only the cases where we impose either Dirichlet or Neumann conditions on all $d$ free bosons, even though we have also established this result for a number of other boundary conditions.

\subsection{Evaluating the cosmological constant}
\label{sec:4.1}

To move in the moduli space of the Narain compactification, we have to use the exactly marginal vertex operator
\be
\V^\epsilon_{1,1}(z,\bar z)=\epsilon_{ij} \p X^i(z)\bar{\p} X^j(\bar z)\,,
\ee
where setting $\alpha^\prime=1$, we have the OPE
$$
\p X^i(z)\p X^j(0)=-\frac{g^{ij}}{2}\frac{1}{z^2}+{\rm reg}\,.
$$
Now suppose we have a D$p$-brane, characterized by a gluing automorphism
\be
\p X_i(z)=\Omega_i^{\;j}\, \bar{\p} X_j(\bar z), \quad z=\bar z\,,
\ee
where the gluing tensor $\Omega$ satisfies the compatibility relation
\begin{align}
\Omega g\Omega^T=g\,.\label{eq:comp}
\end{align}
Applying the doubling trick, the cosmological constant \eqref{cosmo-pert} will be then computed using the bilocal holomorphic operator
\be
\tilde{\V}^\epsilon_{1,1}(z,z^\ast) =\epsilon_{ik}(\Omega^{T})^k_{\;\,j}\, \p X^i(z)  \p X^j( z^*)\equiv \tilde\epsilon_{ij}\,   \p X^i(z)  \p X^j( z^*)\,,
\ee
where we define $\tilde\epsilon= \epsilon\Omega^{T}$.
Defining $V^\epsilon= c\bar{c}\V^\epsilon_{1,1}$ and denoting by $\Lambda_k$ the contribution to the cosmological constant \eqref{cosmo-pert} at order $\mathcal{O}(\epsilon^k)$, we have
\begin{subequations}
	\begin{align}
	g_s^{(0)}\Lambda_1 &=-\omega_\text{c}(V^\epsilon,l_{0,0})\,,\\
	g_s^{(0)}\Lambda_2
	&=-\frac12\left(\omega_\text{c}\!\left(l_{0,0}\,,\frac{b_0^+}{L_0^+}\bar{P}_0^+ l_2(V^\epsilon,V^\epsilon)\right)+\omega_\text{o}\!\left(m_{1,0}(V^\epsilon),\frac{b_0}{L_0}\bar{P}_0m_{1,0}(V^\epsilon)\right)\right)\,,
	\end{align}
\end{subequations}
where in the last line we have focused on the choice of local coordinates discussed in \ref{sec:3.4}, where $l_{1,0}=0$.
Starting with the first order deformation, we have
\begin{subequations}
\begin{align}
g_s^{(0)}\Lambda_1&=\frac{1}{4\pi^2}\langle V^\epsilon |c_0^-\|B(E)\rangle\!\rangle\\
&=-\frac{1}{4\pi^2}\langle\V_{1,1}^\epsilon\|B(E)\rangle\!\rangle_\mathrm{matter}\,,
\end{align}
\end{subequations}
where the minus sign comes from the ghost part of the overlap. Converting this to a correlator on the UHP and applying the doubling trick, we then obtain
\begin{subequations}
\begin{align}
g_s^{(0)}\Lambda_1 &=-\frac{1}{4\pi^2}\langle\V^\epsilon_{1,1}(0,0)\rangle_{\rm disk}\\
&=-\frac{1}{\pi^2}\langle\V_{1,1}^\epsilon(i,\bar i)\rangle_{\rm UHP}\\
&=-\frac{1}{\pi^2}\tilde \epsilon_{ij}\langle \p X^i(i)\p X^j(-i)\rangle_{\mathbb{C}}\\
&=-\frac{1}{8\pi^2} \tilde\epsilon_{ij}g^{ij}\langle 0\|B(E)\rangle\!\rangle\,,
\end{align}
\end{subequations}
where the matter correlator is normalized such that the one-point function of the identity operator gives the boundary state $g$-function $\langle 0\|B(E)\rangle\!\rangle$. That is, we can write\footnote{Here we understand $\mathrm{tr}[\ldots]$ to be the pure matrix trace, that is, we explicitly keep track the factor $g^{-1}$ to emphasize the dependence on the Narain moduli.}
\begin{align}
\Lambda_1&=-\frac{1}{2\pi^2g_s^{(0)}}\langle 0\|B(E)\rangle\!\rangle\times\bigg(+\frac{1}{4}\tr[\tilde\epsilon g^{-1}]\bigg)\,.
\end{align}
To compute the second order deformation, we use \eqref{amplfin} to find
\begin{subequations}
\begin{align}
g_s^{(0)}\Lambda_2&=\frac{1}{8\pi^2}\langle 0\|B(E)\rangle\!\rangle\left(\int_0^{1/{\lo}^{2}} ds \,s^{\eo}+\int_{1/{\lo}^{2}}^1 ds \,y(s)^{\ec}\right)(1-s^2)\times\nonumber\\
&\hspace{3cm}\times\tilde\epsilon_{ij}\tilde\epsilon_{kl}\left(\frac{g^{ij}g^{kl}}{16 s^2}+\frac{g^{ik}g^{jl}}{(1-s)^4}+\frac{g^{il}g^{jk}}{(1+s)^4}\right){\Bigg |}_{\footnotesize\hspace{-1.7mm}\begin{array}{c}\ec\to 0\\ \eo\to 0\end{array}}\\
&=\frac{1}{8\pi^2}\langle 0\|B(E)\rangle\!\rangle\left(\int_0^{1/{\lo}^{2}} ds \,s^{\eo}+\int_{1/{\lo}^{2}}^1 ds \,y(s)^{\ec}\right)(1-s^2)\times\nonumber\\
&\hspace{3cm}\times\left(\frac{(\tr[\tilde\epsilon g^{-1}])^2}{16 s^2}+\frac{\tr[\tilde\epsilon g^{-1}\tilde\epsilon^Tg^{-1}]}{(1-s)^4}+\frac{\tr[\tilde\epsilon g^{-1}\tilde\epsilon g^{-1}]}{(1+s)^4}\right){\Bigg |}_{\footnotesize\hspace{-1.7mm}\begin{array}{c}\ec\to 0\\ \eo\to 0\end{array}}\,.
\end{align}
\end{subequations}
The regularization is now completely parallel to \cite{Maccaferri:2021ksp} and gives the finite result
\begin{align}
\Lambda_2=-\frac{1}{2\pi^2g_s^{(0)}}\langle 0\|B(E)\rangle\!\rangle\bigg(\frac1{32} (\tr[\tilde\epsilon g^{-1}])^2+ \frac1{32}\tr[\tilde\epsilon g^{-1}\tilde\epsilon^Tg^{-1}]-\frac{1}{16}\tr[\tilde\epsilon g^{-1}\tilde\epsilon g^{-1}]\bigg)\,.
\end{align}
Notice that it does not depend on the open string stub parameter $\lo$, as it should. 
In particular it is not difficult to realize that for a single compact free boson (radius deformation) we recover the vanishing result reported in \cite{Maccaferri:2021ksp}.
Altogether, we therefore obtain that the total cosmological constant $\Lambda(\epsilon)$ gives the shifted disk partition function
\begin{align}
-\frac{1}{2\pi^2g_s^{(0)}}\langle 0\|B(E)\rangle\!\rangle+\Lambda(\epsilon)&=-\frac{1}{2\pi^2g_s^{(0)}}\langle 0\|B(E)\rangle\!\rangle\bigg(1+\frac{1}{4}\tr[\tilde\epsilon g^{-1}]+\frac1{32} (\tr[\tilde\epsilon g^{-1}])^2+\nonumber\\
&\hspace{1.7cm}+ \frac1{32}\tr[\tilde\epsilon g^{-1}\tilde\epsilon^Tg^{-1}]-\frac{1}{16}\tr[\tilde\epsilon g^{-1}\tilde\epsilon g^{-1}]+\mathcal{O}(\epsilon^3)\bigg)\,.\label{eq:Zshift}
\end{align}
This result is valid for any D-brane system on the Narain lattice which is specified by the gluing automorphism $\Omega$. Now it is time to verify our conjecture \eqref{conj} by comparing the r.h.s.\ of \eqref{eq:Zshift} with the deformation of the disk partition function computed directly from the boundary state for various D$p$-branes in the sigma-model picture.

\subsection{Comparison with the $g$-function deformation}
\label{sec:4.2}
As is usually the case, the SFT parameter $\epsilon$ is not directly identified with the deformation parameter $\epsilon_\sigma$ in the $\sigma$-model. The relation can be however obtained order by order by comparing the spectrum of physical fluctuations around the classical solution $\Phi(\epsilon)$ with the spectrum of closed string primaries of the Narain compactification with moduli $E+\epsilon_\sigma$. In appendix \ref{app:C} we have performed this analysis in pure closed string field theory in the limit $\lc\to\infty$, following the computation of Sen appearing in \cite{Sen:2019jpm}. The result is\footnote{A similar relation has been discussed in \cite{Hull:2009zb} and \cite{Michishita:2006dr}. We thank B.\ Zwiebach for pointing this out to us.}
\begin{align}
\epsilon(\epsilon_\sigma) = \epsilon_\sigma -\frac{1}{2}\epsilon_\sigma g^{-1}\epsilon_\sigma +\mathcal{O}(\epsilon_\sigma^{3})\,.\label{eq:eesigmaB}
\end{align}
With this important information, let us manipulate \eqref{eq:Zshift} by substituting $\tilde\epsilon = \epsilon\Omega^T$ 
\begin{align}
&-\frac{1}{2\pi^2g_s^{(0)}}\langle 0\|B(E)\rangle\!\rangle+\Lambda(\epsilon(\epsilon_\sigma))=\nonumber\\
&\hspace{0.7cm}=-\frac{1}{2\pi^2g_s^{(0)}}\langle 0\|B(E)\rangle\!\rangle\bigg(1+\frac{1}{4}\tr[\epsilon_\sigma\Omega^T g^{-1}]
-\frac{1}{8}\tr[\epsilon_\sigma g^{-1}\epsilon_\sigma \Omega^T g^{-1}]+\nonumber\\
&\hspace{1.1cm}+\frac1{32} (\tr[\epsilon_\sigma\Omega^T g^{-1}])^2+ \frac1{32}\tr[\epsilon_\sigma g^{-1} \epsilon_\sigma^T g^{-1}]-\frac{1}{16}\tr[\epsilon_\sigma\Omega^T g^{-1}\epsilon_\sigma\Omega^T g^{-1}]\bigg)+\mathcal{O}(\epsilon_\sigma^3)\,,\label{eq:Zshiftsub}
\end{align}
where we have used the compatibility relation \eqref{eq:comp} to note that the $\mathrm{tr}[\epsilon\epsilon^T]$-term is independent of the gluing automorphism $\Omega$.
We will now evaluate the r.h.s.\ of \eqref{eq:Zshiftsub} for the specific cases, when we impose pure Dirichlet and pure Neumann BCs along the Narain lattice and check that it factorizes into the expected shift in the $g$-function times a universal factor (i.e.\ independent of the choice of conformal BCs) which we interpret as a deformation of the closed string coupling constant $g_\mathrm{s}^{(0)}$.

\subsubsection{Pure Dirichlet BCs}

For the pure Dirichlet BCs, we have to set $\tensor{(\Omega^{\mathrm{D}})}{_i^j} = -\tensor{\delta}{_i^j}$. From the boundary state $\|B^{\mathrm{D}}(E)\rangle\!\rangle$, we can also extract the $g$-function
\begin{align}
\langle 0\|B^{\mathrm{D}}(E)\rangle\!\rangle = 2^{-\frac{d}{4}}(\mathrm{det}\, g)^{-\frac{1}{4}}\,.
\end{align}
Deforming the Narain modulus as $E\longrightarrow E^\ast= E+\epsilon_\sigma$, the metric changes as
\begin{align}
g\longrightarrow g^\ast = g+\frac{1}{2}(\epsilon_\sigma+\epsilon_\sigma^T)\,.
\end{align}
Hence, we can calculate the corresponding deformed $g$-function as
\begin{subequations}
\begin{align}
\langle 0\|B^{\mathrm{D}}(E_\ast)\rangle\!\rangle &= 2^{-\frac{d}{4}}(\mathrm{det}\, g^\ast)^{-\frac{1}{4}}\\
&= 2^{-\frac{d}{4}}(\mathrm{det}\, g)^{-\frac{1}{4}} e^{-\frac{1}{4}\mathrm{tr}\log [\delta+\frac{1}{2}g^{-1}(\epsilon_\sigma+\epsilon_\sigma^T)]} \\[0.7mm]
&=\langle 0\|B^{\mathrm{D}}(E)\rangle\!\rangle\, e^{-\frac{1}{8}\mathrm{tr}[g^{-1}(\epsilon_\sigma+\epsilon_\sigma^T)]+\frac{1}{32}\mathrm{tr}[g^{-1}(\epsilon_\sigma+\epsilon_\sigma^T)g^{-1}(\epsilon_\sigma+\epsilon_\sigma^T)]+\mathcal{O}(\epsilon_\sigma^3)}\,,
\end{align}
\end{subequations}
where in the last step, we have Taylor-expanded the logarithm in the exponent. Expanding also the exponential, we obtain
\begin{align}
\langle 0\|B^{\mathrm{D}}(E+\epsilon_\sigma)\rangle\!\rangle &=\langle 0\|B^{\mathrm{D}}(E)\rangle\!\rangle\,\bigg(1-\frac{1}{8}\mathrm{tr}[g^{-1}(\epsilon_\sigma+\epsilon_\sigma^T)]+\frac{1}{128}(\mathrm{tr}[g^{-1}(\epsilon_\sigma+\epsilon_\sigma^T)])^2+\nonumber\\
&\hspace{3.7cm}+\frac{1}{32}\mathrm{tr}[g^{-1}(\epsilon_\sigma+\epsilon_\sigma^T)g^{-1}(\epsilon_\sigma+\epsilon_\sigma^T)]+\mathcal{O}(\epsilon_\sigma^3)
\bigg)\,.
\end{align}
On the other hand, substituting for the specific $\Omega$ into the r.h.s.\ of the SFT calculation \eqref{eq:Zshiftsub}, we eventually obtain
\begin{align}
&-\frac{1}{2\pi^2g_s^{(0)}}\langle 0\|B^{\mathrm{D}}(E)\rangle\!\rangle+\Lambda^{\mathrm{D}}(\epsilon(\epsilon_\sigma))=\nonumber\\
&\hspace{1cm}=-\frac{1}{2\pi^2}\frac{1}{g_s^{(0)}}\bigg(1-\frac1{32}\tr[\epsilon_\sigma g^{-1} \epsilon_\sigma^T g^{-1}]+\mathcal{O}(\epsilon_\sigma^3)\bigg)\times\nonumber\\
&\hspace{2.4cm}\times \langle 0\|B^{\mathrm{D}}(E)\rangle\!\rangle
\bigg(1-\frac{1}{8}\tr[(\epsilon_\sigma +\epsilon_\sigma ^T)g^{-1}]+\frac1{128} (\tr[(\epsilon_\sigma +\epsilon_\sigma ^T) g^{-1}])^2+\nonumber\\
&\hspace{6cm}+\frac{1}{32}\mathrm{tr}[g^{-1}(\epsilon_\sigma+\epsilon_\sigma^T)g^{-1}(\epsilon_\sigma+\epsilon_\sigma^T)]+\mathcal{O}(\epsilon_\sigma^3)\bigg)\,.\label{eq:deffinal}
\end{align}
Hence, provided that we are able to identify the deformed string coupling constant $g_s^{(*)}(\epsilon_\sigma)$ as
\begin{align}
\frac{1}{g_s^{(*)}(\epsilon_\sigma)}=\frac{1}{g_s^{(0)}}\bigg(1-\frac1{32}\tr[\epsilon_\sigma g^{-1} \epsilon_\sigma^T g^{-1}]+\mathcal{O}(\epsilon_\sigma^3)\bigg)
=\frac{1}{g_s^{(0)}}\bigg(1-\frac1{32}\tr[\epsilon g^{-1} \epsilon^T g^{-1}]+\mathcal{O}(\epsilon^3)\bigg)
\,,\label{eq:gdef}
\end{align}
we can rewrite \eqref{eq:deffinal} as
\begin{align}
-\frac{1}{2\pi^2g_s^{(0)}}\langle 0\|B^{\mathrm{D}}(E)\rangle\!\rangle+\Lambda^{\mathrm{D}}(\epsilon(\epsilon_\sigma))=-\frac{1}{2\pi^2g_s^{(*)}(\epsilon_\sigma)}\langle 0\|B^{\mathrm{D}}(E+\epsilon_\sigma)\rangle\!\rangle\,,
\end{align}
which verifies our conjecture \eqref{conj} in the case of pure Dirichlet BCs. In order to see that \eqref{eq:gdef} indeed gives the correct deformation of $g_s$, one would ideally have to compute an independent observable, which directly measures the value of $g_s$. In the absence of such an observable, we will check the plausibility of \eqref{eq:gdef} indirectly by verifying that it is indeed a universal function of the Narain modulus (i.e.\ that it is independent of the particular D-brane system we consider for the calculation of the cosmological constant). This we will do by redoing the calculation for other examples of conformal boundary conditions on the Narain lattice.

\subsubsection{Pure Neumann BCs}

The pure Neumann BCs on a generic Narain lattice are given by the gluing automorphism
\begin{align}
\tensor{(\Omega^\mathrm{N})}{_i^j}=\tensor{[(\delta+Bg^{-1})(\delta-Bg^{-1})^{-1}]}{_i^j}\,,
\end{align}
while the $g$-function can be calculated from the corresponding boundary state $\|B^\mathrm{N}(E)\rangle\!\rangle$. One obtains
\begin{align}
\langle 0\|B^\mathrm{N}(E)\rangle\!\rangle = 2^{-\frac{d}{4}}[\mathrm{det}\,(g-Bg^{-1}B)]^{\frac{1}{4}}\,.
\end{align}
More conveniently, we can express $\langle 0\|B^\mathrm{N}(E)\rangle\!\rangle$ in the DBI-like form as a multiple of the pure Dirichlet $g$-function as
\begin{align}
\langle 0\|B^\mathrm{N}(E)\rangle\!\rangle = \sqrt{\mathrm{det}\,E}\,\langle 0\|B^\mathrm{D}(E)\rangle\!\rangle\,.
\end{align}
Upon deforming the Narain modulus by $E\longrightarrow E^\ast = E+\epsilon_\sigma$, we can calculate the corresponding deformed $g$-function. Performing analogous algebra as in the case of the pure Dirichlet conditions, we obtain
	\begin{align}
&\langle 0\|B^\mathrm{N}(E+\epsilon_\sigma)\rangle\!\rangle=\nonumber\\
&\hspace{1cm}=\langle 0\|B^\mathrm{N}(E)\rangle\!\rangle
\bigg(1-\frac{1}{8}\mathrm{tr}[g^{-1}(\epsilon_\sigma+\epsilon_\sigma^T)]+\frac{1}{128}(\mathrm{tr}[g^{-1}(\epsilon_\sigma+\epsilon_\sigma^T)])^2+\nonumber\\
	&\hspace{4cm}+\frac{1}{32}\mathrm{tr}[g^{-1}(\epsilon_\sigma+\epsilon_\sigma^T)g^{-1}(\epsilon_\sigma+\epsilon_\sigma^T)]+\mathcal{O}(\epsilon_\sigma^3)
	\bigg)\times\nonumber\\
	&\hspace{3cm}\times \bigg(1+\frac{1}{2}\mathrm{tr}[g^{-1}(\delta+Bg^{-1})^{-1}\epsilon_\sigma]+\frac{1}{8}(\mathrm{tr}[g^{-1}(\delta+Bg^{-1})^{-1}\epsilon_\sigma])^2+\nonumber\\
	&\hspace{4cm}-\frac{1}{4}\mathrm{tr}[g^{-1}(\delta+Bg^{-1})^{-1}\epsilon_\sigma g^{-1}(\delta+Bg^{-1})^{-1}\epsilon_\sigma]+\mathcal{O}(\epsilon_\sigma^3)\bigg)\label{eq:BN}
	\end{align}
Starting with the calculation \eqref{eq:Zshift} of the disk partition function shifted by the SFT cosmological constant evaluated for the pure Neumann boundary conditions, we have
\begin{align}
&-\frac{1}{2\pi^2g_s^{(0)}}\langle 0\|B^{\mathrm{N}}(E)\rangle\!\rangle+\Lambda^{\mathrm{N}}(\epsilon)
=\nonumber\\
&\hspace{1cm}=-\frac{1}{2\pi^2g_s^{(0)}}\langle 0\|B^\mathrm{N}(E)\rangle\!\rangle \bigg(1
+\frac{1}{4}\mathrm{tr}\big[ g^{-1}\epsilon (\Omega^{\mathrm{N}})^T\big]-\frac{1}{16}\mathrm{tr}\big[ g^{-1}\epsilon (\Omega^{\mathrm{N}})^T g^{-1}\epsilon (\Omega^{\mathrm{N}})^T\big]+\nonumber\\
&\hspace{4.9cm}
+\frac{1}{32}\big(\mathrm{tr}\big[ g^{-1}\epsilon (\Omega^{\mathrm{N}})^T\big]\big)^2
+\frac{1}{32}\mathrm{tr}\big[g^{-1}\epsilon g^{-1}\epsilon^T\big]
+\mathcal{O}(\epsilon^3)\bigg)\,.\label{eq:step1}
\end{align}
Realizing that we can rewrite
\begin{subequations}
	\begin{align}
(\Omega^{\mathrm{N}})^T g^{-1}&=(\delta+g^{-1}B)^{-1} (\delta-g^{-1}B)g^{-1}\\
	&=g^{-1}(\delta+Bg^{-1})^{-1}(\delta-Bg^{-1})\\
	&=	g^{-1}[2(\delta+Bg^{-1})^{-1}-\delta]\,,
	\end{align}
\end{subequations}
and substituting for $\epsilon$ in terms of the sigma-model modulus $\epsilon_\sigma$, we can factorize the r.h.s.\ of \eqref{eq:step1} as
	\begin{align}
&-\frac{1}{2\pi^2g_s^{(0)}}\langle 0\|B^{\mathrm{N}}(E)\rangle\!\rangle+\Lambda^{\mathrm{N}}(\epsilon(\epsilon_\sigma))=\nonumber\\
&\hspace{1cm}	=-\frac{1}{2\pi^2g_s^{(0)}}\langle 0\|B^\mathrm{N}(E)\rangle\!\rangle
	\bigg(1-\frac1{32}\tr[\epsilon_\sigma g^{-1} \epsilon_\sigma^T g^{-1}]+\mathcal{O}(\epsilon_\sigma^3)\bigg)\times\nonumber\\
	&\hspace{4.3cm}\times
	\bigg(1-\frac{1}{4}\mathrm{tr}[g^{-1}\epsilon_\sigma]+\frac{1}{32}(\mathrm{tr}[g^{-1}\epsilon_\sigma])^2+\nonumber\\
	&\hspace{4.9cm}+\frac{1}{16}\mathrm{tr}[g^{-1}\epsilon_\sigma g^{-1}\epsilon_\sigma]
	+\frac{1}{16}\mathrm{tr}[g^{-1}\epsilon_\sigma g^{-1}\epsilon_\sigma^T]
	+\mathcal{O}(\epsilon_\sigma^3)
	\bigg)\times\nonumber\\
	&\hspace{2.3cm}\times \bigg(1+\frac{1}{2}\mathrm{tr}[g^{-1}(\delta+Bg^{-1})^{-1}\epsilon_\sigma]+\frac{1}{8}(\mathrm{tr}[g^{-1}(\delta+Bg^{-1})^{-1}\epsilon_\sigma])^2+\nonumber\\
	&\hspace{2.9cm}-\frac{1}{4}\mathrm{tr}[g^{-1}(\delta+Bg^{-1})^{-1}\epsilon_\sigma g^{-1}(\delta+Bg^{-1})^{-1}\epsilon_\sigma]+\mathcal{O}(\epsilon_\sigma^3)\bigg)
\,,
	\end{align}
that is, comparing with \eqref{eq:BN},
\begin{align}
&-\frac{1}{2\pi^2g_s^{(0)}}\langle 0\|B^{\mathrm{N}}(E)\rangle\!\rangle+\Lambda^{\mathrm{N}}(\epsilon(\epsilon_\sigma)) =\nonumber\\
&\hspace{3cm}= -\frac{1}{2\pi^2g_s^{(0)}}
\bigg(1-\frac1{32}\tr[\epsilon_\sigma g^{-1} \epsilon_\sigma^T g^{-1}]+\mathcal{O}(\epsilon_\sigma^3)\bigg)\langle 0\|B^\mathrm{N}(E+\epsilon_\sigma)\rangle\!\rangle\,.
\end{align}
We observe that the pure Neumann disk partition function shifted by the SFT cosmological constant can be identified with the shift in the pure Neumann $g$-function times the same factor, which was also present for the pure Dirichlet boundary conditions (and which is indeed present for all other examples of boundary conditions which we have checked, such as mixed Dirichlet-Neumann). Hence, identifying this factor with the BC-independent shift in the string coupling constant as in \eqref{eq:gdef}, we obtain that our conjecture \eqref{conj} holds in the form
\begin{align}
-\frac{1}{2\pi^2g_s^{(0)}}\langle 0\|B^{\mathrm{N}}(E)\rangle\!\rangle+\Lambda^{\mathrm{N}}(\epsilon(\epsilon_\sigma))=-\frac{1}{2\pi^2g_s^{(\ast)}(\epsilon_\sigma)}\langle 0\|B^\mathrm{N}(E+\epsilon_\sigma)\rangle\!\rangle \,.
\end{align}

\section{Conclusions}\label{sec:5}

In this paper we have presented a complete classical picture of the vacuum energy that is generated in bosonic string theory when we deform the closed string background in the presence of D-branes and we have identified such a vacuum energy with the shift in the disk partition function before and after the background shift.\footnote{We do not consider possible contributions from the pure closed SFT action, which is independent of the D-brane background  and it has been recently discussed in \cite{Erler:2022agw}.} This provides a generalization of Sen's conjecture, where the closed string background was fixed.

 This work gives a more rigorous and more regular approach than our previous investigations involving Witten OSFT coupled to the Ellwood invariant \cite{Maccaferri:2021ksp}. At a technical level, thanks to the framework developed in this paper, we can now systematically deal with infrared divergences coming from open and closed string degenerations and very soft (or even zero-momentum) open-closed  amplitudes can be rigorously defined and possibly computed. Although open-closed string field theory is more complicated than Witten OSFT (both technically and conceptually) we have found that, at the end of the day, many of the relevant computations that are doable in Witten OSFT \cite{Maccaferri:2021ksp} are still doable here,  with the advantage that closed string degenerations are now unambiguosly tamed. It is reasonable to expect that  progress in the construction of SFT vertices \cite{Firat:2021ukc, Cho:2019anu, Moosavian:2019pmd, Costello:2019fuh, Headrick:2018ncs}
 will further improve this situation.  
 
 At a more formal level our analysis gives a concrete  framework to approach the full open-closed background independence, when we want to deal with generic closed string background shifts.  Our results can also be useful  in the context of purely closed string field theory where it is not easy to define non-trivial and non-vanishing gauge-invariant observables \cite{Erler:2022agw}. Our cosmological constant can provide gauge invariant information on the change in the closed string background,  which essentially means to use matter (D-branes)  to probe purely `gravitational' phenomena. 

A generic observation that is possible to derive from our conjecture \eqref{conj} and from the detailed example of section \ref{sec:4} is that a  closed string field theory solution in general  changes the string coupling constant, not just  the CFT background.   To first order in the closed string deformation it is known that the main field which is responsible to change the string coupling  is the ghost dilaton \cite{Bergman:1994qq, Belopolsky:1995vi} but, already at second order, a purely matter bulk deformation will also (generically) shift the string coupling, even if the ghost dilaton is not switched on. We saw this effect explicitly in section \ref{sec:4} through the  $\tr[\epsilon\cdot\epsilon^T]$  contribution in the cosmological constant, which is independent of the D-brane system and which `factor out' after matching the SFT deformation parameter with the sigma-model one, following appendices \ref{app:B} and  \ref{app:C}. We think that this is an interesting point that deserves further study and we are indeed working at the moment at generalizing our results to deformations involving the ghost-dilaton \cite{progress1}. Ideally, it would be desirable to have a SFT observable that independently measures the shift in the coupling constant which is induced by a classical solution.

Another natural continuation of this work is the extension to superstring field theory. Recently Kunitomo  has constructed the  OCHA for the  RNS superstring  \cite{Kunitomo:2022qqp}. However, just as in the bosonic string, this is not sufficient to define a gauge-invariant cosmological constant: we also need purely closed string couplings on the disk to construct the full SDHA which is the basic algebraic structure at the heart of this matter. We will report on this construction as well as on the associated ${\cal N}=2$ localization \cite{Maccaferri:2021ulf, Erbin:2019spp, Maccaferri:2019ogq, Maccaferri:2018vwo} in the context of Type II theories in \cite{progress2}.

After having addressed these extensions to our present results, it will be time to confront  with the intrinsic quantum nature of open-closed string field theory. We expect that explorations in this direction will improve our understanding of the non-perturbative structure of string theory.

\section*{Acknowledgments}
We would like to thank Ted Erler for discussions and Ashoke Sen and Barton Zwiebach for reading the manuscript and providing comments. We particularly acknowledge Ashoke Sen for suggesting the use of $\mathrm{SL}(2,\mathbb{C})$ vertices in explicit computations. JV also thanks INFN Turin for their hospitality during the initial stages of this work. The work of JV was supported by the NCCR SwissMAP that
is funded by the Swiss National Science Foundation.
The work of CM is partially supported by the MIUR PRIN
Contract 2015MP2CX4 ``Non-perturbative Aspects Of Gauge Theories And Strings''.

\appendix


\section{$\mathrm{SL}(2,\mathbb{C})$ vertices for the the disk two-point function}\label{app:A}
In this appendix we explicitly build the OC-vertices that are needed for the bulk two-point function on the disk, which (beside the open and closed  BRST operators) are $\left(l_2, l_{0,0}, l_{1,0}, m_{1,0}\right)$. We will find it useful to use $\mathrm{SL}(2,\mathbb{C})$ vertices \footnote{We thank Ashoke Sen for suggesting this.}, so that full off-shell computations will be possible without having to deal with complicated conformal maps. Our strategy will be to write down a reasonable ansatz involving $\mathrm{SL}(2,\mathbb{C})$ local coordinates and then to constrain the free parameters so that the basic homotopy relation 
\be
l_2l_{0,0}+[Q_\text{c},l_{1,0}]+l_{0,1}m_{1,0}=0\,,\label{hom2}
\ee
will be satisfied when evaluated between two generic closed string states
\be
\omega_\text{c}\left(\Phi_1,l_2(l_{0,0},\Phi_2)+[Q_\text{c},l_{1,0}](\Phi_2)+l_{0,1}m_{1,0}(\Phi_2)     \right)=0\,, \quad \forall \Phi_1,\Phi_2 \in {\cal H}^\text{c}\,,
\ee
where we note that in order for the symplectic form to be non-zero, the degrees $d(\Phi_1)$, $d(\Phi_2)$ of $\Phi_1$ and $\Phi_2$ need to satisfy $d(\Phi_1)+d(\Phi_2)\in 2\mathbb{Z}+1$.
In fact, as we are going to show, it is possible to find an explicit $\mathrm{SL}(2,\mathbb{C})$ solution to the homotopy relation \eqref{hom2} where $l_{1,0}=0$, that is 
\be
l_2l_{0,0}+l_{0,1}m_{1,0}=0\,.\label {homshort}
\ee
As explained in the main text, this means that it is possible to construct the bulk two-point disk amplitude with just two Feynman diagrams: two open-closed vertices connected by an open string propagator and two closed strings interacting, connected to the boundary state by a closed string propagator. Let us then start with the first term in \eqref{homshort} and evaluate it between two generic closed string states 
\begin{align}
\omega_\text{c}\left(\Phi_1,l_2(l_{0,0},\Phi_2)\right)&=-\wc(l_{0,0},l_2(\Phi_1,\Phi_2))\,.
\end{align}
We define the closed string product $l_2$ as
\be
\ket{l_2(\Phi_1,\Phi_2)}=b_0^-\delta(L_0^-)\,f_1\circ\Phi_1(0,\bar 0)\,f_2\circ\Phi_2(0,\bar 0)\,\ket0_{\mathrm{SL}(2,\mathbb{C})}\,,
\ee
where the $\mathrm{SL}(2,\mathbb{C})$ functions $f_{1,2}$ are defined from  three $\mathrm{SL}(2,\mathbb{C})$ maps
\begin{subequations}
\begin{align}
g_1(w)&=\frac32\,\frac{w-\lc}{3\lc+w}\,,\\
g_2(w)&=\frac32\,\frac{\lc-w}{3\lc+w}\,,\\
g_3(w)&=\frac32 \,\frac{\lc}{w}\,,
\end{align}
\end{subequations}
where $\lc>1$ is a tunable parameter corresponding to a closed string stub created by the action of $\lc^{-L_0^+}$ in the local coordinate $w$. With these maps, we can define the basic cubic closed string vertex as\footnote{The closed string BPZ conjugation is performed with $I_c(z)=\frac1z$}
\begin{subequations}
\begin{align}
&(-1)^{d(\Phi_3)}\wc(\Phi_3,l_2(\Phi_1,\Phi_2))=\nonumber\\
&\hspace{2cm}=\aver{g_3\circ\Phi_3(0,\bar 0)g_1\circ\Phi_1(0,\bar 0)g_2\circ\Phi_2(0,\bar 0)}_{\mathbb{C}}\\
&\hspace{2cm}=\aver{I_c\circ\Phi_3(0,\bar 0)\,I_c\circ g_3^{-1}\circ g_1\circ\Phi_1(0,\bar 0)\,I_c\circ g_3^{-1}\circ g_2\circ\Phi_2(0,\bar 0)}_{\mathbb{C}}\,,
\end{align}
\end{subequations}
from which we can read off 
\begin{align}
f_1(w)&=\frac1\lc\,\frac{w-\lc}{3\lc+w}=-f_2(w)\,.
\end{align}
Notice that thanks to the use of $\mathrm{SL}(2,\mathbb{C})$ maps, correlators are always computed on the complex plane (because $f(\mathbb{C})=\mathbb{C}$ for $f\in \mathrm{SL}(2,\mathbb{C})$). By construction this vertex obeys $[Q_\text{c},l_2]=0$. 

The next object we have to define is the zero-closed-string product $l_{0,0}$. In generality we take it to be  
the boundary state of the starting open string background BCFT$_0$, $\| B_0\rangle\!\rangle$, with a tunable closed string stub attached
\be
l_{0,0}=\frac{1}{(2\pi i)^2}\lb^{-L_0^+}\| B_0\rangle\!\rangle=-\frac{1}{4\pi^2}b_0^-c_0^-\delta_{L_0^-}\lb^{-L_0^+}\| B_0\rangle\!\rangle\,,
\ee
where we used that the boundary state obeys the gluing condition $b_0^-\| B_0\rangle\!\rangle=0$ and it is BRST invariant. Moreover, the prefactor $(2\pi i)^{-2}$ ensures correct normalization of 1-point disk amplitudes. Consequently, we have
\be
\wc(l_{0,0},\Phi)=+\frac{1}{4\pi^2}\aver{\left(\frac w\lb \right)\circ c_0^{-}\Phi(w,\bar w){\Big|}_{w=0}}_{\rm disk}\,.
\ee
We are now ready to write down the first part of the homotopy relation \eqref{homshort} which, after mapping the disk to the UHP by $z_{\rm u}=i\frac{1-z_{\rm d}}{1+z_{\rm d}}$, reads
\begin{subequations}
\begin{align}
\wc\left(\Phi_1,l_2(l_{0,0},\Phi_2)\right)&=-\wc(l_{0,0},l_2(\Phi_1,\Phi_2))\\
&=-\frac{1}{4\pi^2}\aver{\frac1\lb\circ f_1\circ\Phi_1(0,\bar 0)\,\frac1\lb\circ f_2\circ\Phi_2(0,\bar 0)}_{\rm disk}\\
&=-\frac{1}{4\pi^2}\aver {I_o\circ\eta\circ\Phi_1(0,\bar 0)\,\eta\circ\Phi_2(0,\bar 0)}_{\rm UHP}\,,
\end{align}
\end{subequations}
where
\begin{align}
\eta(w)=&\,i\,\left(\frac{3\lb\lc-1}{3\lb\lc+1}\right)\,\frac{1+\frac{\lb\lc+1}{\lc(3\lb\lc-1)}w}{1+\frac{\lb\lc-1}{\lc(3\lb\lc+1)}w}\,.
\end{align}
Let us now consider the second term in the homotopy relation \eqref{homshort}, namely
\begin{align}
\wc(\Phi_1,l_{0,1}m_{1,0}(\Phi_2))=-(-1)^{d_1}\wo(m_{1,0}(\Phi_1),m_{1,0}(\Phi_2))=\aver{m_{1,0}(\Phi_1),m_{1,0}(\Phi_2)}_\text{o}\,.
\end{align}
To compute this quantity, we define the $\mathrm{SL}(2,\mathbb{C})$ open-closed vertex $m_{1,0}$
\begin{align}
(-1)^{d(\Phi_1)} \ket{m_{1,0}(\Phi_1)}=\frac{1}{2\pi i }\Big[\widetilde{m\circ \Phi_1(0,\bar 0)}\Big]\ket0_{\mathrm{SL}(2,\mathbb{R})}\,,\label{ocv}
\end{align}
where $\widetilde{\cdots}$ means that we use the doubling trick to express the anti-holomorphic part of the vertex operator $\Phi(z,\bar z)$ as an holomorphic insertion on the lower half plane \footnote{This operation depends on the boundary conditions of BCFT$_0$.}.
Again, we start with a basic open-closed overlap
\begin{align}
(-1)^{d(\Psi)+d(\Phi)}\wo(\Psi,m_{1,0}(\Phi))=\frac{1}{2\pi i }\aver{f_o\circ\Psi(0)\,f_c\circ\Phi(0,\bar 0)}_{\rm UHP}\,,
\end{align}
where we introduce maps
\begin{subequations}
\begin{align}
f_o(w)&=\frac w\lo\,, \\
f_c(w)&=i\,\frac{1+\frac{w}{\beta_1}}{1+\frac{w}{\beta_2}}\,,
\end{align}
\end{subequations}
and where the open string stub $\lo>1$ and $\beta_{1,2}>1$ are tunable parameters which will be related to the closed string parameters $\lc,\lb$ in order to solve \eqref{homshort}. A short comment is in order: this is a well-defined vertex if the open string  patch $f_o(w),\, (|w|\leq 1)$ does not intersect  the closed string patch $f_c(w),\, (|w|\leq 1)$. Choosing for definiteness $\beta_2>\beta_1$ it is easy to check that this implies the constraint
\begin{align}
\lo>\frac{\beta_1}{\beta_2}\,\frac{1-\beta_2}{1-\beta_1}\,, \quad (\beta_2>\beta_1)\,.
\end{align}
To get the function $m(w)$ in \eqref{ocv} we observe\footnote{The open string BPZ conjugation is performed with $I_o(w)=-\frac1w$.}
\begin{subequations}
\begin{align}
\frac{1}{2\pi i }\aver{f_o\circ\Psi(0)\,f_c\circ\Phi(0,\bar 0)}_{\rm UHP}&=\frac{1}{2\pi i }\aver{I_o\circ\Psi(0)\,I_o\circ f_o^{-1}\circ f_c\circ\Phi(0,\bar 0)}_{\rm UHP}\\
&=(-1)^{d(\Psi)+d(\Phi)}\omega_\mathrm{o}({\Psi,m_{1,0}(\Phi)})\,,
\end{align}
\end{subequations}
which fixes
\be
m(w)=\frac i\lo\,\frac{1+\frac w{\beta_2}}{1+\frac w{\beta_1}}\,.
\ee
From here we easily get 
\begin{subequations}
\begin{align}
\wc(\Phi_1,l_{0,1}m_{1,0}(\Phi_2))&=-(-1)^{d(\Phi_1)+d(\Phi_2)}\frac{1}{4\pi^2}\aver{I_o\circ m\circ\Phi_1(0,\bar 0)\,m\circ\Phi_2(0,\bar 0)}_{\rm UHP}\,,\\
\wc\left(\Phi_1,l_2(l_{0,0},\Phi_2)\right)&=-\frac{1}{4\pi^2}\aver {I_o\circ\eta\circ\Phi_1(0,\bar 0)\,\eta\circ\Phi_2(0,\bar 0)}_{\rm UHP}\,.
\end{align}
\end{subequations}
Hence, recalling that $d(\Phi_1)+d(\Phi_2)\in 2\mathbb{Z}+1$, we conclude that \eqref{homshort} is satisfied if and only if
\be
\eta(w)=m(w)\,,
\ee
which means that we need to put
\begin{subequations}
\begin{align}
\lo=&\frac{3\lb\lc+1}{3\lb\lc-1}\,,\\
\beta_2=&\frac{3\lb\lc-1}{\lb\lc+1}\,\lc\,,\\
\beta_1=&\frac{3\lb\lc+1}{\lb\lc-1}\,\lc\,.
\end{align}
\end{subequations}
To simplify a bit, we can choose the stub of $l_{0,0}$ ($\lb$) to coincide with he stubs of $l_2$ ($\lc$) and in this case we get
\begin{subequations}
\begin{align}
\lo=&\frac{3\lc^2+1}{3\lc^2-1}\,,\\
\beta_2=&\frac{3\lc^2-1}{\lc^2+1}\,\lc\,,\quad\quad (\lb=\lc)\,,\\
\beta_1=&\frac{3\lc^2+1}{\lc^2-1}\,\lc\,.
\end{align}
\end{subequations}

\section{Mass-shift in string field theory}\label{app:B}

In this appendix we include general discussion of the BRST cohomology deformation which is induced by a classical SFT solution describing a marginally deformed background. For the sake of concreteness, we will consider marginal deformations in closed SFT and the associated shifts in the masses of the closed-string physical states. For constructing the elements of the deformed cohomology, we will follow the perturbative method described in \cite{Sen:2019jpm}, emphasizing the fact that the microscopic cohomology can be expressed as an $L_\infty$ homotopy transfer of an effective cohomology. Since the construction is be purely algebraic, it is applicable also for calculating mass-shift of open-string physical states arising due to marginal deformations in open SFT (which is governed by an $A_\infty$ algebra). 


\subsection{Marginal deformations}

Let us fix a background where the matter $\mathrm{CFT}_0$ can be factorized as
\begin{align}
\mathrm{CFT}_0 = \mathrm{CFT}_0^{\mathbb{R}_{1,D}}\oplus \mathrm{CFT}_0^{M}\,,
\end{align}
where $\mathrm{CFT}_0^{\mathbb{R}_{1,D}}$ is the $c=D+1$ CFT of $D+1$ free bosons $X^0\equiv Y,X^1,\ldots,X^D$ propagating in $\mathbb{R}_{1,D}$ (for $D\geq 0$) and $\mathrm{CFT}_0^{M}$ is a $c=25-D$ CFT with discrete spectrum of conformal dimensions (and which is otherwise arbitrary) describing a compactification. The boson $Y$ propagates along the time-like direction of $\mathbb{R}_{1,D}$. We will consider an exactly marginal deformation which is induced by a weight $(1,1)$ matter field $\mathbb{V}_{1,1}\in \mathrm{CFT}_0^{M}$ and which thus only deforms the compactification $M$. Writing as before $V = c\bar{c}\mathbb{V}_{1,1}\in\mathrm{ker}\, L_0^+$ and using the notation of \cite{Erbin:2020eyc}, we can write the most general form of the closed SFT classical solution $\Phi_\mu$ describing the marginal deformation as
\begin{align}
\Phi_\mu = \pi_1  \tilde{\mathbf{I}}\, e^{\wedge \phi_\mu}\,,
\end{align}
where the string field 
\begin{align}
\phi_\mu = \sum_{k=1}^\infty \mu^k \phi_k
\end{align}
(with $\phi_1\equiv V$) belongs to $\mathrm{ker}\,L_0^+$ and satisfies the effective closed SFT equation of motion
\begin{align}
0=\pi_1 \tilde{\mathbf{l}}\, e^{\wedge \phi_\mu} =\sum_{j=1}^\infty \frac{1}{j!}\tilde{l}_{j}(\phi_\mu^{\wedge j})\,.\label{eq:EffEOM}
\end{align}
Here $\tilde{\mathbf{I}}$ denotes the canonical inclusion of $SP_0^+\mathcal{H}_c$ into $S\mathcal{H}_c$ deformed by interactions (where $S\mathcal{H}_c$ denotes the symmetrized tensor coalgebra on $\mathcal{H}_\mathrm{c}$). This can be defined in terms of a contracting homotopy operator $h$, where in Siegel gauge, we would usually have $h = (b_0^+/L_0^+)(1-P_0^+)$ where $P_0^+$ is the projector onto $\mathrm{ker}\,L_0^+$. The products $\tilde{l}_k$ would then be given by the corresponding homotopy transfer of the microscopic $L_\infty$ products $l_k$. For example, up to quadratic order in $\mu$, we could extract
\begin{align}
\Phi_\mu = \mu \phi_1  -\frac{1}{2!}\mu^2 \frac{b_0^+}{L_0^+}(1-P_0^+)l_2(\phi_1,\phi_1)+\mu^2 \phi_2+\mathcal{O}(\mu^3)\,,
\end{align}
where $\phi_1\equiv V,\phi_2, \ldots$ satisfy
\begin{subequations}
	\begin{align}
	0&= Q\phi_1\,,\\
	0&= Q\phi_2 +\frac{1}{2!}P_0^+ l_2(\phi_1,\phi_1)\,,\\
	&\hspace{0.2cm}\vdots\nonumber
	\end{align}
\end{subequations}
The solution for the marginal deformation considered in the main body of this paper therefore corresponds to the special situation where we can set $\phi_2 = \phi_3 =\ldots = 0$. 

In order to facilitate the description of the associated deformation of the BRST cohomology, we will need to introduce the projector $P_\mu^+$ onto the subset
\begin{align}
\mathcal{K}_\mu = \{\Phi\in\mathcal{H}_0 : L_0\Phi = \mathcal{O}(\mu)\}\,.
\end{align}
Clearly we have $\mathrm{ker}\,L_0\subset\mathcal{K}_\mu \subset \mathcal{H}_c$. Denoting by $P_0^+$ the projector onto $\mathrm{ker}\, L_0^+$, we have $	P_0^+ P_\mu^+= P_\mu^+	P_0^+  = P_0^+$ and also $[Q,P_0^+]=0=[Q,P_\mu^+]$.
Note that since the solution $\phi_\mu$ is defined to live in $\mathrm{ker}\,L_0^+$ and its matter part excites only fields in $\mathrm{CFT}_0^M$ which has discrete spectrum, we can actually replace $P_0^+$ with $P_\mu^+$ with no effect on the above derivation. However, it will make a difference whether we use $P_0^+$ or $P_\mu^+$ when we start inserting states from $\mathcal{K}_\mu$ which do not belong to $\mathrm{ker}\,L_0^+$. This will be neccessary below when we attempt to define an observable yielding the mass-shifts of physical states. Hence, from now on it will be more convenient to consider the homotopy transfer with respect to the contracting homotopy operator $(b_0^+/L_0^+)(1-P_\mu^+)$.

\subsection{Shifted cohomology}
 
Let us now consider the cohomology of the effective BRST operator around the deformed closed-string background given by the classical effective solution $\phi_\mu\in\mathrm{ker}\, L_0$. Following the strategy of \cite{Sen:2019jpm}, we will expect that for a given element $u_0\in \mathrm{ker}\,L_0$ of the original (undeformed) cohomology around $\mathrm{CFT}_0$ (thus satisfying $Qu_0=0$), one may hope to be able extend this to a continuous family of elements $u_\mu\in\mathcal{K}_\mu$, which are closed in the deformed effective cohomology, namely, which satisfy the condition
\begin{align}
0=\pi_1 \tilde{\mathbf{l} }\, (e^{\wedge \phi_\mu} \wedge u_\mu)=\sum_{j=0}^\infty \frac{1}{j!}\tilde{l}_{j+1}(\phi_\mu^{\wedge j}\wedge u_\mu)\,.\label{eq:DefCohomEff}
\end{align}
Note that using the effective equation of motion \eqref{eq:EffEOM} satisfied by $\phi_\mu$, the $\pi_1$ projector in front can actually be dropped and we remain with the condition
\begin{align}
\tilde{\mathbf{l} }\, (e^{\wedge \phi_\mu} \wedge u_\mu)=0\,.
\end{align}
Also note that in order to make perturbative sense of the l.h.s.\ of \eqref{eq:DefCohomEff}, it is important to consider the homotopy transfer $\tilde{l}_k$ of the microscopic products $l_k$ with respect to the projector $P_\mu^+$ on the enlarged kernel $\mathcal{K}_\mu$ of $L_0$ so as to protect against acting with $1/L_0^+$ on states with conformal dimensions of $\mathcal{O}(\mu)$.
The corresponding elements $U_\mu$ of the cohomology with respect to the BRST operator shifted by the full microscopic classical solution $\Phi_\mu$ (where $U_0 = u_0$) can then be computed in terms of $u_\mu$ by applying homotopy transfer as
\begin{align}
e^{\wedge \Phi_\mu} \wedge U_\mu =  \tilde{\mathbf{I}}\,   (e^{\wedge \phi_\mu} \wedge u_\mu)\,.
\end{align}
The states $U_\mu$ then indeed satisfy the appropriate condition
\begin{align}
\sum_{j=0}^\infty \frac{1}{j!}{l}_{j+1}(\Phi_\mu^{\wedge j}\wedge U_\mu)=\pi_1 \mathbf{l}\, (e^{\wedge \Phi_\mu} \wedge U_\mu) = \pi_1 \mathbf{l}\tilde{\mathbf{I}}\, (e^{\wedge \phi_\mu} \wedge u_\mu)= \pi_1\mathbf{I}\tilde{\mathbf{l}}\, (e^{\wedge \phi_\mu} \wedge u_\mu)=0\,.
\end{align}
on the shifted microscopic BRST operator (where in the third equality, we have used the chain-map property $\mathbf{l}\tilde{\mathbf{I}}=\mathbf{I}\tilde{\mathbf{l}}$). In particular, let us focus on elements $u_0$ of the undeformed cohomology, which can be expressed as
\begin{align}
u_0 = c\bar{c}e^{ik_0 Y} \mathbb{U}_{h,\bar{h}}\in\mathrm{ker}\,L_0\,,
\end{align}
where $\mathbb{U}_{h,\bar{h}}\in \mathrm{CFT}_M$ is a dimension $(h,\bar{h})$ matter primary operator (with $h=\bar{h}$ to saturate the level-matching condition) and where the undeformed energy $k_0$ of the state $u_0$ satisfies the mass-shell constraint (denoting $h_\mathrm{tot}=h+\bar{h}$)
\begin{align}
-2+h_\mathrm{tot}-\frac{k_0^2}{2}=0\,,
\end{align}
so that $Qu_0=0$. Let us now attempt to find a solution $u_\mu$ to the condition \eqref{eq:DefCohomEff} in the form
\begin{align}
u_\mu =  c\bar{c}e^{ik_\mu Y} \mathbb{U}_{h,\bar{h}}\in\mathcal{K}_\mu\,,
\end{align}
where $k_\mu = k_0+\mathcal{O}(\mu)$.
Provided that such a solution exists, it is therefore completely determined by the function $k_\mu$. On the other hand, given $u_\mu$, we can extract the corresponding value of $k_\mu$ by calculating (recall that we assume $h=\bar{h}$)
\begin{align}
Qu_\mu = \frac{1}{2}\Big(\!-\!2+h_\mathrm{tot}-\frac{k_\mu^2}{2}\Big)c\bar{c}(\p c+\bar{\p}\bar{c}) e^{ik_\mu Y} \mathbb{U}_{h,\bar{h}}\,.
\end{align}
Assuming then that the matter fields $\mathbb{U}_{h,\bar{h}}$ have normalized 2-point functions, we can extract
\begin{align}
\frac{k_\mu^2}{2}=2\omega_c(\tilde{u}_\mu,Qu_\mu)-2+h_\mathrm{tot}\,,\label{eq:kfromu}
\end{align}
where we have introduced the undeformed cohomology element $\tilde{u}_\mu =  c\bar{c}e^{-ik_\mu Y} \mathbb{U}_{h,\bar{h}}$ which is dual to $u_\mu$. Let us try to solve for $k_\mu$ perturbatively by introducing the approximate solutions $k_\mu^{(r)}$ which satisfy
\begin{align}
k_\mu=k_\mu^{(r)}+\mathcal{O}(\mu^{r+1})\,.
\end{align}
Correspondingly, we can then introduce approximate cohomology elements
\begin{align}
u_\mu^{(r)} =  c\bar{c}e^{ik_\mu^{(r)} Y} \mathbb{U}_{h,\bar{h}}\in\mathcal{K}_\mu\,,
\end{align}
which are only exact up to order $\mu^r$, that is $u_\mu = u_\mu^{(r)} +\mathcal{O}(\mu^{r+1})$. Going in the opposite direction, we could again extract $k_\mu^{(r)}$ from $u_\mu^{(r)}$ by a calculation analogous to \eqref{eq:kfromu}.
The approximate states $u_\mu^{(r)}$ then satisfy the truncated conditions
\begin{align}
\sum_{j=0}^r \frac{1}{j!}\tilde{l}_{j+1}((\phi_\mu^{(r+1-j)})^{\wedge j}\wedge u_\mu^{(r-j)})=0\,,\label{eq:DefCohomEffApprox}
\end{align}
for all $r\geq 0$, where we have also introduced the truncations 
\begin{align}
\phi_\mu^{(r)} = \sum_{k=1}^r \mu^k \phi_k\,,
\end{align}
of the effective classical solution $\phi_\mu$, which satisfy the truncated equations of motion
\begin{align}
\sum_{j=1}^r \frac{1}{j!} \tilde{l}_j((\phi_\mu^{(r+1-j)})^{\wedge j})=0\,,\label{eq:EOMApprox}
\end{align}
for all $r\geq 1$.
More explicitly, for $r=0,1,2,\ldots$, we obtain equations
\begin{subequations}
	\begin{align}
	0&=Qu_\mu^{(0)} \,,	\\[1mm]
	0&=Qu_\mu^{(1)} +\tilde{l}_2(\phi_\mu^{(1)}, u_\mu^{(0)})\,,	\label{eq:cond1}\\
	0&=Qu_\mu^{(2)} +\tilde{l}_2(\phi_\mu^{(2)}, u_\mu^{(1)})
	+\frac{1}{2!}\tilde{l}_3((\phi_\mu^{(1)})^{\wedge 2}, u_\mu^{(0)})
	\,,	\label{eq:cond2}\\
	&\hspace{0.2cm}\vdots\nonumber
	\end{align}
\end{subequations}
Hence, starting with some undeformed element $u_\mu^{(0)}\equiv u_0$, we can calculate $k_\mu^{(1)}$ by considering the correlator
\begin{align}
-\frac{1}{2}\Delta_\mu^{(1)} \equiv \omega_c(\tilde{u}_\mu^{(0)},\tilde{l}_2(\phi_\mu^{(1)},u_\mu^{(0)})) \,,\label{eq:Delta1}
\end{align}
(where on the r.h.s.\ we only have known quantities) since using \eqref{eq:cond1}, the quantity $\Delta_\mu^{(1)}$ is then equal to
$2\omega_c(\tilde{u}_\mu^{(1)},Qu_\mu^{(1)})$. This can then be substituted into \eqref{eq:kfromu} to obtain $k_\mu^{(1)}$. Knowing $k_\mu^{(1)}$ (and therefore $u_\mu^{(1)}$) we can calculate $k_\mu^{(2)}$ from 
\begin{align}
-\frac{1}{2}\Delta_\mu^{(2)} \equiv\omega_c(\tilde{u}_\mu^{(1)},\tilde{l}_2(\phi_\mu^{(2)},u_\mu^{(1)})) +\frac{1}{2}\omega_\mathrm{c}(\tilde{u}_\mu^{(0)}, \tilde{l}_3((\phi_\mu^{(1)})^{\wedge 2}, u_\mu^{(0)}))\,,\label{eq:Delta2}
\end{align}
because using $\eqref{eq:cond2}$, the quantity $\Delta_\mu^{(2)}$ is equal to $2\omega_c(\tilde{u}_\mu^{(2)},Qu_\mu^{(2)})$, which in turn gives $k_\mu^{(2)}$ through \eqref{eq:kfromu}. Continuing in this manner, we can gradually improve the approximate solutions $k_\mu^{(r)}$ and therefore build (up to arbitrary order in $\mu$) a state $u_\mu$ which satisfies the condition \eqref{eq:DefCohomEff}. Indeed, in general, for arbitrary $r\geq 1$, we can extract $k_\mu^{(r)}$ in terms of $k_\mu^{(r-1)},k_\mu^{(r-2)},\ldots,k_\mu^{(0)}$ by calculating
\begin{align}
-\frac{1}{2}\Delta_\mu^{(r)} \equiv \sum_{j=1}^{r} \frac{1}{j!}\omega_c(\tilde{u}_\mu^{(r-j)},\tilde{l}_{j+1}((\phi_\mu^{(r+1-j)})^{\wedge j}\wedge u_\mu^{(r-j)}))\,,
\end{align}
because, using the condition \eqref{eq:DefCohomEffApprox}, $\Delta_\mu^{(r)}$ is equal to $2\omega_c(\tilde{u}_\mu^{(r)},Qu_\mu^{(r)})$.
Furthermore, according to \cite{Sen:2019jpm}, $\Delta_\mu^{(r)}$ can be identified 
with the $\mathcal{O}(\mu^r)$-approximation of the shift in the total conformal dimension $h_\mathrm{tot}=h+\bar{h}$ of the state $\mathbb{U}_{h,\bar{h}}$ as we deform the  matter background $\mathrm{CFT}_0^M$ by turning on a vev for $\mathbb{V}_{1,1}$. 

\section{Relating the SFT and sigma-model deformation parameters}\label{app:C}

In this appendix we will apply the method of calculating the mass-shift using SFT in the special case where we take $\mathrm{CFT}_0^M$ to be the theory of $c=d$ free bosons $X^1,\ldots,X^d$ compactified to a $d$-dimensional Narain lattice, whose moduli are encoded in the metric $g_{ij}$ and the Kalb-Ramond field $B_{ij}$. The corresponding closed SFT marginal deformation will be turned on by the matter state $\mathbb{V}_{1,1}=\epsilon_{ij}\p X^i\bar{\p}X^j$. By calculating the SFT mass-shift of arbitrary momentum and winding modes along the compactified directions and comparing with the known formula for their conformal weights in terms of the Narain moduli, we will then establish (up to quadratic order) the relation between the SFT modulus $\epsilon$ and the corresponding parameter $\epsilon_\sigma$ which controls the marginal deformation in the sigma-model description.

\subsection{Narain spectrum and sigma-model deformations}

Given a Narain lattice $(g,B)$, the corresponding spectrum of momentum and winding modes is spanned by the vertex operators
\begin{align}
\mathbb{V}_{k_\mathrm{L}(E),k_\mathrm{R}(E)}(z,\bar{z})=e^{i(k_\mathrm{L}(E))_i (X_\mathrm{L})^i+i(k_\mathrm{R}(E))_i (X_\mathrm{R})^i}(z,\bar{z})\,,
\end{align}
where, denoting $E_{ij}=g_{ij}+B_{ij}$, we have 
\begin{subequations}
	\begin{align}
(k_\mathrm{L}(E))_i &= k_i + E_{ij}w^j\,,\\
(k_\mathrm{R}(E))_i &= k_i - E^T_{ij}w^j\,,
	\end{align}
\end{subequations}
with $k_i\in \mathbb{Z}$, $w^j\in \mathbb{Z}$ arbitrary integers. The corresponding left and right conformal dimensions can be computed as
\begin{subequations}
	\label{eq:weights}
\begin{align}
h_\mathrm{L}(E)&=\frac{1}{4}(k_\mathrm{L}(E))^T g^{-1}k_\mathrm{L}(E)\,,\\
h_\mathrm{R}(E)&=\frac{1}{4}(k_\mathrm{R}(E))^T g^{-1}k_\mathrm{R}(E)\,,
\end{align}
\end{subequations}
where $g^{ij} =\frac{1}{2}(E+E^T)^{ij}$ are the elements of the inverse metric. The total conformal weight is then given simply as $h_\mathrm{tot} = h_\mathrm{L}+h_\mathrm{R}$. We now want to deform the Narain lattice by varying 
\begin{align}
E\longrightarrow E^\prime = E+\epsilon_\sigma
\end{align}
for some deformation $\epsilon_\sigma$, whose symmetric part deforms the metric $g$, while its antisymmetric part deforms the Kalb-Ramond field $B$. Calculating the corresponding deformation of the left- and right-moving momenta, we obtain
\begin{subequations}
\begin{align}
(k_\mathrm{L}(E+\epsilon_\sigma))_i &= (k_\mathrm{L}(E))_i+ (\epsilon_\sigma)_{ij}w^j\,,\\
(k_\mathrm{R}(E+\epsilon_\sigma))_i &= (k_\mathrm{R}(E))_i- (\epsilon_\sigma^T)_{ij}w^j\,.
\end{align}
\end{subequations}
Substituting these into \eqref{eq:weights} and expanding the deformation of the inverse metric, we obtain the deformed total conformal weight up to second order in $\epsilon_\sigma$ as
\begin{align}
h_\mathrm{tot}(E+\epsilon_\sigma) &= h_\mathrm{tot}(E)+\nonumber\\
&\hspace{1cm}-\frac{1}{2}(k_\mathrm{L}(E))^T g^{-1}\epsilon_\sigma g^{-1}k_\mathrm{R}(E)
+\nonumber\\
&\hspace{1cm}
+\frac{1}{4}(k_\mathrm{L}(E))^T g^{-1}\epsilon_\sigma g^{-1}\epsilon_\sigma g^{-1}k_\mathrm{R}(E)+\nonumber\\
&\hspace{3cm}
+\frac{1}{8}(k_\mathrm{L}(E))^T g^{-1}\epsilon_\sigma g^{-1}\epsilon_\sigma^T g^{-1}k_\mathrm{L}(E)
+\nonumber\\
&\hspace{3cm}
+\frac{1}{8}(k_\mathrm{R}(E))^T g^{-1}\epsilon_\sigma^T g^{-1}\epsilon_\sigma g^{-1}k_\mathrm{R}(E)+\nonumber\\
&\hspace{1cm}
+\mathcal{O}(\epsilon_\sigma^3)\,.\label{eq:CFTshift}
\end{align}

\subsection{SFT mass-shift}

Let us now compare the shift \eqref{eq:CFTshift} in the Narain spectrum to the corresponding mass-shift calculated to second order using closed SFT. We will therefore consider a marginal deformation which is excited by the massless state
\begin{align}
V = \epsilon_{ij} c\bar{c} \p X^i \bar{\p}X^j\,,
\end{align}
where the comparison of the SFT and CFT mass-shifts will yield $\epsilon$ as a function of $\epsilon_\sigma$. Calculating the first-order approximation of the mass-shift $\Delta_\epsilon^{(1)}$ from \eqref{eq:Delta1}, we obtain
\begin{subequations}
\begin{align}
\Delta_\epsilon^{(1)}&=-2\epsilon_{ij}\,\omega_c(c\bar{c}\mathbb{V}_{-k_\mathrm{L},-k_\mathrm{R}}e^{-ik_0Y},{l}_2(c\bar{c} \p X^i \bar{\p}X^j,c\bar{c}\mathbb{V}_{k_\mathrm{L},k_\mathrm{R}}e^{ik_0Y}))\\
&=-\frac{1}{2}(k_\mathrm{L}(E))^T g^{-1}\epsilon g^{-1}k_\mathrm{R}(E)\,.
\end{align}
\end{subequations}
Comparing this with \eqref{eq:CFTshift}, we conclude that up to linear order we can identify $\epsilon = \epsilon_\sigma +\mathcal{O}(\epsilon_\sigma^2)$. At the same time, this gives
\begin{align}
\frac{(k_\epsilon^{(1)})^2}{2}=h_\mathrm{tot}(E)-2-\frac{1}{2}(k_\mathrm{L}(E))^T g^{-1}\epsilon g^{-1}k_\mathrm{R}(E)\,.
\end{align}
In order to improve the approximation to the quadratic order in $\epsilon$, we have to carefully evaluate
\begin{align}
\Delta_\epsilon^{(2)}&=2\epsilon_{ij	}\omega_c(
c\bar{c}\mathbb{V}_{-k_\mathrm{L},-k_\mathrm{R}}e^{-ik_\epsilon^{(1)} Y}
,{l}_2(c\bar{c} \p X^i \bar{\p}X^j,c\bar{c}\mathbb{V}_{k_\mathrm{L},k_\mathrm{R}}e^{ik_\epsilon^{(1)} Y}))+\nonumber\\[3mm]
&\hspace{0.4cm}
 +\epsilon_{ij}\epsilon_{kl}\omega_\mathrm{c}(c\bar{c}\mathbb{V}_{-k_\mathrm{L},-k_\mathrm{R}}e^{-ik_0Y}, {l}_3(c\bar{c} \p X^i \bar{\p}X^j,c\bar{c} \p X^k \bar{\p}X^l, c\bar{c}\mathbb{V}_{k_\mathrm{L},k_\mathrm{R}}e^{ik_0Y}))+\nonumber\\[2mm]
&\hspace{0.4cm} -\epsilon_{ij}\epsilon_{kl}\omega_\mathrm{c}(c\bar{c}\mathbb{V}_{-k_\mathrm{L},-k_\mathrm{R}}e^{-ik_0Y}, {l}_2(\frac{b_0^+}{L_0^+}\bar{P}_0^+l_2(c\bar{c} \p X^i \bar{\p}X^j,c\bar{c} \p X^k \bar{\p}X^l), c\bar{c}\mathbb{V}_{k_\mathrm{L},k_\mathrm{R}}e^{ik_0Y}))+\nonumber\\
&\hspace{0.4cm}  -2\epsilon_{ij}\epsilon_{kl}\omega_\mathrm{c}(c\bar{c}\mathbb{V}_{-k_\mathrm{L},-k_\mathrm{R}}e^{-ik_0Y}, {l}_2(c\bar{c} \p X^i \bar{\p}X^j,\frac{b_0^+}{L_0^+}\bar{P}_0^+l_2(c\bar{c} \p X^k \bar{\p}X^l, c\bar{c}\mathbb{V}_{k_\mathrm{L},k_\mathrm{R}}e^{ik_0Y})))\,.
\end{align}
The individual terms in this expression can be computed using the results of \cite{Sen:2019jpm}, since our calculation is just a $d>1$ generalization of the mass-shift calculated in Section 7.1 of \cite{Sen:2019jpm}. The calculation can be performed independently of the local coordinates in the limit of large closed-string stub $\lc$, one obtains
\begin{align}
\Delta_\epsilon^{(2)} &= -\frac{1}{2}(k_\mathrm{L}(E))^T g^{-1}\epsilon g^{-1}k_\mathrm{R}(E)
+\frac{1}{8}(k_\mathrm{L}(E))^T g^{-1}\epsilon g^{-1}\epsilon^T g^{-1}k_\mathrm{L}(E)
+\nonumber\\
&\hspace{7cm}
+\frac{1}{8}(k_\mathrm{R}(E))^T g^{-1}\epsilon^T g^{-1}\epsilon g^{-1}k_\mathrm{R}(E)\,.\label{eq:SFTshift}
\end{align}
Hence, comparing \eqref{eq:CFTshift} with \eqref{eq:SFTshift}, it follows that the SFT deformation parameter $\epsilon$ needs to be related to the sigma-model parameter $\epsilon_\sigma$ by the field redefinition
\begin{align}
\epsilon = \epsilon_\sigma -\frac{1}{2}\epsilon_\sigma g^{-1}\epsilon_\sigma +\mathcal{O}(\epsilon_\sigma^{3})\,.\label{eq:eesigma}
\end{align}

\endgroup
\end{document}